\documentclass[10pt]{JHEP3}

\usepackage[centertags]{amsmath} 
\usepackage{amssymb}

\newcommand{\be}{\begin{equation}}
\newcommand{\ee}{\end{equation}}
\newcommand{\ben}{\begin{eqnarray}}
\newcommand{\een}{\end{eqnarray}}
\newcommand{\ba}{\begin{eqnarray}}
\newcommand{\ea}{\end{eqnarray}}
\newcommand{\nn}{\nonumber \\}

\newcommand{\beq}{\begin{equation}}
\newcommand{\eeq}{\end{equation}}

\newcommand{\p}{\partial}
\newcommand{\bi}{\begin{itemize}}
\newcommand{\ei}{\end{itemize}}

\newcommand{\labell}[1]{\label{#1}} 

\newcommand{\mn}{\mu \nu}

\newcommand{\bc}{\begin{center}}
\newcommand{\ec}{\end{center}}

\newcommand{\stone}{{\text{ST1}}}

\newcommand{\stfive}{{\text{ST5}}}

\newcommand{\vtwo}{{\text{V2}}}
\newcommand{\vthre}{{\text{V3}}}
\newcommand{\vfour}{{\text{V4}}}
\newcommand{\vfive}{{\text{V5}}}
\newcommand{\vtone}{{\text{VT1}}}
\newcommand{\vttwo}{{\text{VT2}}}
\newcommand{\vtthre}{{\text{VT3}}}
\newcommand{\vtfour}{{\text{VT4}}}
\newcommand{\vtfive}{{\text{VT5}}}
\newcommand{\qvone}{{\text{QV1}}}
\newcommand{\qvtwo}{{\text{QV2}}}
\newcommand{\qvthre}{{\text{QV3}}}
\newcommand{\qvfour}{{\text{QV4}}}

\title{Hydrodynamics from charged black branes}

\author {Nabamita Banerjee$^{a}$, Jyotirmoy Bhattacharya$^{b}$,
Sayantani Bhattacharyya$^{b}$, Suvankar Dutta$^{a}$, \linebreak R. Loganayagam$^{b}$, and
P. Sur\'owka$^{c, d}$.\\
$^a$Harish-Chandra Research Institute, Chhatnag Road, Jhunsi, Allahabad-211019.\\
$^b$Dept. of Theoretical Physics, Tata Institute of Fundamental Research, Homi Bhabha Rd, \\
Mumbai 400005, India. \\
$^c$Department of Physics, University of Washington, Seattle,  WA 98195-1560, USA. \\
$^d$Institute of Physics, Jagiellonian University, Reymonta 4, 30-059 Krak\'ow, Poland. \\
E-mail:\ \ {\bf nabamita@mri.ernet.in, jyotirmoy@theory.tifr.res.in, sayanta@theory.tifr.res.in,
suvankar@mri.ernet.in, nayagam@theory.tifr.res.in, surowka@u.washington.edu}}

\abstract{We extend the recent work on fluid-gravity correspondence to
charged black-branes by determining the metric duals to arbitrary
charged fluid configuration up to second order in the boundary
derivative expansion. We also derive the energy-momentum tensor and
the charge current for these configurations up to second order in the
boundary derivative expansion. We find a new term in the charge
current when there is a bulk Chern-Simons interaction thus resolving
an earlier discrepancy between thermodynamics of charged rotating
black holes and boundary hydrodynamics. We have also confirmed that
all our expressions are covariant under boundary Weyl-transformations
as expected.}


\preprint{TIFR/TH/08-37}

\begin{document}
{\vskip 1cm} 

\section{Introduction} \label{intro}

Modern theoretical physics provides mathematically precise
descriptions of a bewildering variety of phenomena. The diversity of
the phenomena studies by physicists encourages specialization and a
consequent divergence in the field. We find it satisfying, however,
that this push towards divergence is partially counterbalanced by
periodic theoretical discoveries that unify - i.e. discover precise
mathematical connections between - distinct fields of physics.  In
line with this tradition, recent string theory inspired studies of
classical gravitational dynamics have found a precise mathematical
connection between a long distance limit of the Einstein equations of
gravity and the Navier Stokes equations of fluid dynamics. More
specifically, it has recently been demonstrated that a class of long
distance, regular, locally asymptotically $AdS_{d+1}$ solutions to
Einstein's equations with a negative cosmological constant is in one
to one correspondence with solutions to the charge free Navier Stokes
equations in $d$ dimensions
\cite{Bhattacharyya:2008jc,VanRaamsdonk:2008fp,Dutta:2008gf,
Bhattacharyya:2008ji,Haack:2008cp,Bhattacharyya:2008xc,Loganayagam:2008is}
\footnote{There exists a large literature in deriving linearise
  hydrodynamics from AdS/CFT.  See(\cite{Policastro:2001yc} -
  \cite{Baier:2007fj}). There have been some recent work on
  hydrodynamics with higher derivative corrections
  \cite{Buchel:2008ae,Buchel:2008kd}.}.

The connection between the equations of gravity and fluid dynamics, 
described above, was demonstrated 
essentially by use of the method of collective coordinates. The
authors of \cite{Bhattacharyya:2008jc,VanRaamsdonk:2008fp,
Dutta:2008gf,Bhattacharyya:2008ji,Haack:2008cp} noted that there 
exists a $d$ parameter set of exact, asymptotically $AdS_{d+1}$ 
 black brane solutions of the gravity equations parameterized by
temperature and velocity.  They then used the `Goldstone' philosophy
to promote temperatures and velocities to fields. The Navier Stokes
equations turn out to be the effective `chiral Lagrangian
equations' of the temperature and velocity collective fields.

This initially surprising connection between gravity in $d+1$
dimensions and fluid dynamics in $d$ dimensions is beautifully
explained by the AdS/CFT correspondence. Recall that a particular
large $N$ and strong coupling limit of that correspondence relates the
dynamics of a classical gravitational theory (a two derivative theory
of gravity interacting with other fields) on AdS$_{d+1}$ space to the
dynamics of a strongly coupled conformal field theory in $d$ flat
dimensions. Now the dynamics of a conformal field theory, at length
scales long compared to an effective mean free path (more accurately
an equilibration length scale) is expected to be well described by the
Navier Stokes equations. Consequently, the connection between long
wavelength solutions of gravity and the equations of fluid dynamics -
directly derived in \cite{Bhattacharyya:2008jc} - is a natural
prediction of the AdS/CFT correspondence.  Using the AdS/CFT
correspondence, the stress tensor as a function of velocities and
temperatures obtained above from gravity may be interpreted as the fluid
stress tensor of the dual boundary field theory in its
deconfined phase.

Now consider a conformal field theory that has a conserved charge $Q$
in addition to energy and momentum. This is especially an interesting
extension of the hydrodynamics of the uncharged fluids since the
hydrodynamics of many real fluids has a global conserved charge which
is often just the number of particles that make up the fluid. The long
distance dynamics of such a system is expected to be determined by the
augmented Navier Stokes equations; $\nabla_\mu T^{\mu\nu}=0$ together
with $\nabla_\mu J_Q^\mu=0$, where the stress tensor and charge
current are now given as functions of the temperature, velocity and
charge density, expanded to a given order in the derivative
expansion. The bulk dual description of a field theory with a
conserved charge always includes a propagating Maxwell field. 
Consequently the AdS/CFT correspondence suggests asymptotically
$AdS$ long wavelength solutions of appropriate modifications of the
the Einstein Maxwell equation are in one to one correspondence with
solutions of the augmented Navier Stokes equations described above.

This expectation of the previous paragraph also fits well with the
collective coordinate intuition described above. Recall that the
Einstein Maxwell equations have a well known $d+1$ dimensional set of
charged black brane solutions, parameterized by the brane temperature,
charge density and velocity.  It seems plausible that the effective
Goldstone equations, that arise from the promotion of these $d+1$
dimensional parameters to fields, are simply the augmented Navier
Stokes equations.  In this paper we verify the expectations via a
direct analysis of the relevant bulk equations. More concretely, we
generalize the work out in \cite{Bhattacharyya:2008jc} to set up a
perturbative scheme to generate long wavelength solutions of the
Einstein Maxwell equations plus a Chern Simons term (see below for
more details) order by order in the derivative expansion. We also
implement this expansion to second order, and thereby find explicit
expressions for the stress tensor and charge current of our dual fluid
to second order in the derivative expansion.

In this paper we work with the Einstein Maxwell equations augmented by
a Chern Simon's term. This is because the equations of IIB SUGRA on
AdS$_5\times$S$^5$ (which is conjectured to be dual to ${\cal N}=4$
Yang Mills) with the restriction of equal charges for the three
natural Cartans, admit a consistent truncation to this system. Under
this truncation, we get the following action 
\begin{equation}\label{actionIntro:eq}
\begin{split}
S=\frac{1}{16\pi G_5}\int \sqrt{-g_5}
\left[R + 12 - F_{AB}F^{AB} - \frac{4\kappa}{3}\epsilon{}^{LABCD}A_L F_{AB} F_{CD} \right] 
\end{split}
\end{equation}
In the above action the size of the $S_5$  has been set to 1.
The value of the parameter $\kappa$ for ${\cal N}=4$ Yang Mills is
given by $\kappa=1/(2\sqrt{3})$ - however, with a view to other
potential applications we leave $\kappa$ as a free parameter in all
the calculations below. Note in particular that our bulk Lagrangian
reduces to the true Einstein Maxwell system at $\kappa=0$.


Our expressions for the charge current and the stress tensor of the
fluid are complicated, and are listed in detail in subsequent sections.
We would however like to point out an important qualitative feature of our
result. Already at first order, and at nonzero $\kappa$, the charge
current includes a term proportional to
$l^\alpha\equiv\epsilon^{\mu\nu\lambda\alpha}u_\mu\nabla_\nu
u_\lambda$.  The presence of this term in the current resolves an
apparent mismatch between the predictions of fluid dynamics and the
explicit form of charged rotating black holes in IIB supergravity
reported in \cite{Bhattacharyya:2007vs}. Note that due to the
presence of the $\epsilon$ symbol, this term is parity odd. However, when
accompanied by a flip in the R-charge of the brane, its sign remains unchanged.
Consequently, this term is CP symmetric in agreement with the expectations of 
CP symmetry of ${\cal N}=4$ Yang Mills theory.

As we have explained above, the reduction of boundary field theory dynamics
is expected to reduce to field theory dynamics only at long wavelength 
compared to an effective mean free path or equilibration length scale.
All the gravitational constructions of this paper also work only in the 
same limit. It is consequently of interest to know the functional form 
of the equilibration length scale of our conformal fluid as a function 
of intensive fluid parameters. 

In the case of ${\cal N}=4$ Yang Mills, it follows from `t Hooft scaling and 
dimensional analysis that, at large $\lambda$, the effective equilibration 
length scale is given by $l_{mfp}=f(\nu)/T$ where $\nu$ is the dimensionless 
chemical potential conjugate to the conserved charge of the theory 
and $T$ being the associated temperature. Explicit
computation within gravity demonstrates that $f(\nu)$ is of unit order for 
generic values of $\nu$. Consequently, at generic values of $\nu$, all the 
considerations of this paper apply only when all fields vary at distances 
and times that are large compared to the local effective temperature. 
However, as was explained in detail in \cite{Bhattacharyya:2007vs}, the 
charged fluid we study in this paper has an upper bound on $\nu$ at 
$\nu_c$. At this special value of $\nu$, the black brane becomes extremal. 
$f(\nu)$ appears to have a simple zero about $\nu_c$. Consequently it appears
to be possible to scale $\nu$ to $\nu_c$ and $T$  to zero simultaneously
while keeping $l_{mfp}$ (and all other thermodynamic densities) finite. 
Thus it would naively appear that the long distance field theory 
dynamics should be well described by fluid dynamics in this coordinated 
extremal limit. It turns out, however, that the bulk gravitational solutions
described in this paper turn singular in the same limit. We are not completely
sure how to interpret this fact. It would certainly be interesting to 
investigate this further.

\textbf{Note Added :} While this draft was in preparation, we became aware of a
similar work by Erdmenger et.al.\cite{Erdmenger:2008rm}.

\section{Notations and Conventions} \label{convnot}

In this section. we will establish the basic conventions and notations that we will use in the rest of the paper. We start with the five-dimensional action\footnote{We use Latin letters $A,B\in\{r,v,x,y,z\}$ 
to denote the bulk indices and $\mu,\nu \in\{v,x,y,z\}$ to denote the boundary indices.}
\begin{equation}\label{action:eq}
\begin{split}
S=\frac{1}{16\pi G_5}\int \sqrt{-g_5}
\left[R + 12 - F_{AB}F^{AB}-\frac{4\kappa}{3}\epsilon{}^{LABCD}A_L F_{AB} F_{CD} \right] 
\end{split}
\end{equation}
which is a consistent truncation of IIB SUGRA Lagrangian on 
AdS$_5\times$S$^5$ background with a cosmological constant $\Lambda=-6$
and the Chern-Simons parameter $\kappa=1/(2\sqrt{3})$ (See for example, 
\cite{cvetic1,jonson,Chamblin:1999hg,Astefanesei:2008wz,Astefanesei:2007vh,cai,Cai:1998ji,tenauth,sabra+liu,cvetic2,gubser,Banerjee:2007by}).
However, for the sake of generality (and to 
keep track of the effects of the Chern-Simons term), we will work with
an arbitrary value of $\kappa$ in the following. In particular, $\kappa=0$
corresponds to a pure Maxwell theory with no Chern-Simons type interactions.

The field equations corresponding to the above action are
\begin{equation}\label{einmax:eq}
\begin{split}
G_{AB}-6 g_{AB} + 2\left[F_{AC}F^C{}_{B}+\frac{1}{4} g_{AB} F_{CD}F^{CD}\right] &= 0\\
\nabla_B F^{AB} + \kappa \epsilon{}^{ABCDE}F_{BC} F_{DE} &=0
\end{split}
\end{equation}
where $g_{AB}$ is the five-dimensional metric, $G_{AB}$ is the five dimensional Einstein tensor. These equations admit an AdS-Reisner-Nordstr\"{o}m black-brane solution 
\begin{equation}\label{brane:eq}
\begin{split}
ds^2&= -2 u_\mu dx^\mu dr - r^2 V(r,m,q) \ u_\mu u_\nu dx^\mu dx^\nu + r^2 P_{\mu\nu} dx^\mu dx^\nu \\
A &= \frac{\sqrt{3} q}{2 r^2} u_\mu dx^\mu, 
\end{split}
\end{equation}
where
\begin{equation}\label{Vu:eq}
\begin{split}
u_\mu dx^\mu = - dv ;\qquad& V(r,m,q) \equiv 1-\frac{m}{r^4}+\frac{q^2}{r^6};\\
P_{\mu\nu} &\equiv \eta_{\mu\nu} + u_{\mu}u_{\nu},
\end{split}
\end{equation}
with $\eta_{\mu\nu}=\text{diag}(-+++)$ being the Minkowski-metric. 
Following the procedure elucidated in \cite{Bhattacharyya:2008jc}, 
we shall take this flat black-brane metric as our zeroth order metric/gauge 
field ansatz and promote the parameters $u_\mu,m$ and $q$ to slowly varying fields
\footnote{Note that the charge we consider here refers to the Maxwell charge $\int_{\partial S} F_{A B} r^A t^B$ in
the bulk (where $r^A$ and $t^A$ are respectively the unit radial normal and 
future pointing time-like normal to the spatial bounday $\partial S$). In the presence of a Chern-Simons
term in the bulk lagrangian (or alternatively, when the boundary
global charge is anomalous), there are other
notions of charge (like Page charge - see, for example
\cite{Marolf:2000cb} ) which are employed in the literature.
The Page charge in the bulk  would be 
$\int_{\partial S} \left( F_{A B} + 2 \kappa ~~\epsilon_{AB}^{~~~~CDE} A_{C}F_{D E}\right)r^A t^B $ 
in our notation .
These other notions of charge in the bulk mirrors the various possible
notions of a global charge
when it is anomalous in the boundary theory. However, in the rest of 
the paper, we shall not
concern ourselves with these subtleties for the following  reason -
for the  solutions in this paper,
F and hence $A \wedge F$ vanishes when restricted to boundary of AdS. In such
a case, the boundary
anomaly is turned off and the definition of conserved charge in the
boundary is unambiguous
(Maxwell charge and Page charge become equal for this subset of
solutions). In fact, for a
specific value of $\kappa$ , this conserved charge refers  to the
unique R-charge of the
boundary super conformal field theory.}.
 
In the course of our calculations, we will often find it convenient to use
the following `rescaled' variables
\begin{equation}\label{rhoMQ:eq}
\begin{split}
\rho &\equiv \frac{r}{R} ;\qquad M\equiv \frac{m}{R^4} ;\qquad Q\equiv
\frac{q}{R^3} ;\qquad Q^2= M-1 \\
\end{split}
\end{equation}
where $R$ is the radius of the outer horizon,i.e., the largest
positive root of the equation $V=0$. The Hawking temperature, 
chemical potential and the charge density of this black-brane
are given by\footnote{In much of the literature the chemical 
potential $\mu$ is taken to be the potential difference between
the boundary and the horizon. However we have chosen a different 
normalization for $\mu$ (and hence the charge density $n$). we 
shall elaborate on this point in subsection \ref{stresscur1}.}
\begin{equation}
\begin{split}
T\equiv\frac{R}{2\pi}(2-Q^2)\ ,\qquad \mu\equiv\frac{2 \sqrt{3} q}{R^2}=2 \sqrt{3} QR \qquad\text{and}\qquad n\equiv  \frac{\sqrt{3} q}{16 \pi G_5}.
\end{split}
\end{equation}
In terms of the rescaled
variables, the outer and the inner horizon are given by
\[ \rho_+\equiv 1 \qquad \text{and} 
\qquad \rho_-\equiv\left[\left(Q^2+1/4\right)^{1/2}-1/2\right]^{1/2} \] 
and the extremality condition $\rho_+=\rho_-$ corresponds to
$(Q^2=2,M=3)$.  We shall assume the black-branes and the corresponding
fluids to be non-extremal unless otherwise specified - this corresponds
to the regime $0< Q^2 < 2$ or $0 < M< 3$ which we will assume henceforth. 

Using the flat black-brane solutions with slowly varying velocity,
temperature and charge fields, our intention is to systematically
determine the corrections to the metric and the gauge
field in a derivative expansion. More precisely, we expand the metric
and the gauge field in terms of derivatives of velocity, temperature
and charge fields of the fluid as
\begin{equation}
\begin{split}
g_{AB} &= g^{(0)}_{AB}+ g^{(1)}_{AB}+g^{(2)}_{AB}+ \ldots \\
A_{M} &= A^{(0)}_M+A^{(1)}_M+A^{(2)}_M+ \ldots \\
\end{split}
\end{equation}
where $g^{(k)}_{AB}$ and $A^{(k)}_M$ contain the k-th derivatives of
the velocity, temperature and the charge fields with
\begin{equation}\label{zerogA:eq}
\begin{split}
g^{(0)}_{AB} dx^A dx^B&= -2 u_\mu(x) dx^\mu dr - r^2 V(r,m(x),q(x)) \
u_\mu(x) u_\nu(x) dx^\mu dx^\nu + r^2 P_{\mu\nu}(x) dx^\mu dx^\nu \\
A^{(0)}_M dx^M &= \frac{\sqrt{3} q(x)}{2 r^2} u_\mu(x) dx^\mu.
\end{split}
\end{equation}

In order to solve the Einstein-Maxwell-Chern-Simons system of
equations, it is necessary to work in a particular gauge for the
metric and the gauge fields. Following \cite{Bhattacharyya:2008jc}, we
choose our gauge to be
\begin{equation}\label{gauge:eq}
\begin{split}
g_{rr}=0 ;\quad g_{r\mu} \propto u_\mu\ ;\quad A_r = 0 ;\quad
Tr[(g^{(0)})^{-1}g^{(k)}] = 0.
\end{split}
\end{equation}

Further, in order to relate the bulk dynamics to boundary
hydrodynamics, it is useful to parameterise the fluid dynamics in the
boundary in terms of a `fluid velocity' $u_\mu$. In case of
relativistic fluids with conserved charges, there are two widely used
conventions of how the fluid velocity should be defined. In this
paper, we will work with the Landau frame velocity 
where the fluid velocity is defined with reference to the energy
transport. In a more practical sense working in the Landau frame
amounts to taking the unit time-like eigenvector of the energy-momentum
tensor at a point to be the fluid velocity at that
point.

Alternatively, one could work in the `Eckart frame' where the fluid
velocity is defined with reference to the charge transport where the
unit time-like vector along the charged current to be the definition
of fluid velocity. Though the later is often the more natural
convention in the context of charged fluids, we choose to use the
Landau's convention for the ease of comparison with the other
literature. We will leave the conversion to the more natural Eckart
frame to future work.

In the next two sections, we will report in some detail the calculations
leading to the determination of the metric and the gauge field up to 
second order in the derivative expansion. This will enable us to determine 
the boundary stress tensor and charge current up to the second order.

\section{First Order Hydrodynamics}\label{firstord}

In this section, we present the computation of the metric and the gauge
field up to first order in derivative expansion, the derivative being
taken with respect to the boundary coordinates. We choose the boundary
coordinates such that $u^\mu = (1,0,0,0)$ at $x^\mu$. Since our
procedure is ultra local therefore we intend to solve for the metric
and the gauge field at first order about this special
point $x^\mu$. We shall then write the result thus obtained in a
covariant form which will be valid for arbitrary choice of boundary
coordinates.

In order to implement this procedure we require the zeroth order metric
and gauge field expanded up to first order. For this we recall that the
parameters m, q and the velocities ($\beta_i$) are functions of the boundary
coordinates and therefore admit an expansion in terms of the boundary
derivatives. These parameters expanded up to first order is given by
\begin{equation}
\begin{split}
m =& m_0 + x^{\mu} \p_{\mu} m^{(0)} + \dots \\
q =& q_0 + x^{\mu} \p_{\mu} q^{(0)} + \dots \\
\beta_i =& x^{\mu} \p_{\mu} \beta^{(0)}_i + \dots \\
\end{split}
\end{equation}
Here $m^{(i)}$, $q^{(i)}$, $\beta^{(i)}$ refers to the 
i-th order correction to mass, charge and velocities respectively.

The zeroth order metric expanded about $x^\mu$ up to first
order in boundary coordinates is given by
\begin{equation} 
\begin{split}
{ds^{(0)}}^2 &= 2\, dv \, dr - r^2 V^{(0)}(r)\, dv^2 + r^2\, dx_i \, dx^i \\
&- 2\, x^{\mu}\, \partial_{\mu}\beta^{(0)}_i \, dx^i \, dr -
2\, x^\mu \partial_\mu
\beta^{(0)}_i \,r^2(1-V^{(0)}(r))\, dx^i \, dv  \\
&- \left(\frac{-x^{\mu}\partial_{\mu} \, m^{(0)}}{r^2} +
\frac{2 q_0 x^{\mu}\partial_{\mu} \, q^{(0)} }{r^4} \right)\, dv^2,
\end{split}
\end{equation}
where $m_0$ and $q_0$ are related to the mass and charge of the background 
blackbrane respectively and 
$$V^{(0)} = 1-\frac{m_0}{r^4}+\frac{q_0^2}{r^6}.$$

Similarly the zeroth order gauge fields expanded about $x^\mu$ up to first order
is given by
\begin{equation}
A = - \frac{\sqrt{3}}{2} \left[ \left(\frac{q_0 + x^{\mu}\partial_{\mu}\, 
q^{(0)}}{r^2}\right) dv
-\frac{q_0}{r^2} x^{\mu}\partial_{\mu}\, \beta^{(0)}_i \, dx^i \right]
\end{equation}

Since the background black brane metric preserves an $SO(3)$  symmetry
\footnote{Here we are referring to the $SO(3)$ rotational symmetry 
in the boundary spatial coordinates.}, the Einstein-Maxwell equations 
separate into equations in scalar, transverse vector and the symmetric 
traceless transverse tensor sectors. This in turn allows us to solve 
separately for $SO(3)$ scalar, vector and symmetric traceless tensor 
components of the metric and the gauge field.

\subsection{Scalars Of $SO(3)$ at first order}\labell{scal1}

The scalar components of first order metric and gauge field perturbations
($g^{(1)}$ and $A^{(1)}$ respectively) are parameterized by the functions $h_1(r)$,
$k_1(r)$ and $w_1(r)$ as follows \footnote{here $i$ runs over the boundary
spatial coordinates, v is the boundary time coordinate and r is the radial coordinate 
in the bulk}
\begin{equation}
\begin{split}
\sum_{i} g^{(1)}_{ii}(r) &=  3 r^2 h_1(r), \\
g^{(1)}_{vv}(r) &= \frac{k_1(r)}{r^2} \\
g^{(1)}_{vr}(r) &= -\frac{3}{2} h_1(r) \\
A^{(1)}_v(r) &= -\frac{\sqrt{3} w_1(r)}{2 r^2}
\end{split}
\end{equation}
Note that $g^{(1)}_{ii}(r)$ and $g^{(1)}_{vr}(r)$ are related to each other by 
the  gauge choice $Tr[(g^{(0)})^{-1}g^{(1)}] = 0$.

\subsubsection*{\it \bf Constraint equations}\label{scalconsteq1}

We begin by finding the constraint equations that constrain various 
derivatives velocity,temperature and charge that appear in the first 
order scalar sector.The constraint equations are obtained by taking a dot 
of the Einstein and Maxwell equations with the vector dual to the one form $dr$. 
If we denote the Einstein and the Maxwell equations by $E_{AB}=0$ and 
$M_{AB}=0$, then there are three constraint relations.

Two of them come from Einstein equations. They are given by
\be
\label{sclconsE1}
g^{rr}E_{vr} + g^{rv} E_{vv}=0\ ,
\ee
and
\be \label{sclconsE2}
g^{rr}E_{rr} + g^{rv} E_{vr}=0\ ,
\ee
and the third constraint relation comes from Maxwell
equations and is given by
\be
\label{sclconsM1}
g^{rr}M_{r} + g^{rv} M_{v}=0 \ .
\ee

Equation \eqref{sclconsE1} reduces to
\be
\label{seinconst}
  \partial_v m^{(0)} =-\frac{4}{3} m_0 \partial_i \beta^{(0)}_i \ . 
\ee
which is same as the conservation of energy in the boundary at the first order
in the derivative expansion, i.e., the above equation is identical to the constraint
(scalar component of the constraint in this case)
\be \label{1stordcon1}
\p_{\mu}T_{(0)}^{\mn}=0 \ .
\ee
on the allowed boundary data.

The second constraint equation \eqref{sclconsE2} in scalar sector 
implies a relation between $h_1(r)$ and $k_1(r)$.
\be \label{1stordcon2}
2 \partial_i \beta^{(0)}_i r^5 + 12 r^6 h_1(r) + 4 q_0 w_1(r) - m_0
  r^3 h_1'(r) + 3 r^7 h_1'(r) - r^3 k_1'(r) - 2 q_0 r w_1'(r) = 0.
\ee

The constraint relation coming from Maxwell equation
(See Eq.~\eqref{sclconsM1}) gives
\begin{equation}\label{smaxconst}
 \partial_v \ q^{(0)} = - q_0 \ \partial_i \beta^{(0)}_i \ .
\end{equation}
This equation can be interpreted as the conservation of boundary current
density at the first order in  the derivative expansion.
\be
\p_{\mu} J_{(0)}^{\mu} = 0.
\ee

We now proceed to find the scalar part of the metric dual 
to a fluid configuration which obeys the above constraints.

\subsubsection*{\it \bf Dynamical equations and their solutions}\labell{scaldyneq1}

Among the Einstein equations four are $SO(3)$ scalars (namely the $vv$, $rv$, $rr$ 
components and the trace over the boundary spatial part). Further the $r$ and 
$v$-components of the Maxwell equations constitute two other equations in this 
sector. Two specific linear combination of the $rr$ and $vv$ components of the 
Einstein equations constitute the two constraint equations in \eqref{seinconst}. 
Further, a linear combination of the $r$ and $v$-components of the Maxwell equations 
appear as a constraint equation in \eqref{smaxconst}. Now among the six equations 
in the scalar sector we can use any three to solve for the unknown functions 
$h_1(r)$, $k_1(r)$ and $w_1(r)$ and we must make sure that the solution satisfies
the rest. The simplest two equations among these dynamical equations are
\begin{equation}\label{rrein}
 5 h_1'(r)+r h_1''(r) = 0.
\end{equation}
which comes from the $rr$-component of the Einstein equation and
\begin{equation}\label{rmax}
 6 q_0 h_1'(r) + w_1'(r) - r w_1''(r) = 0. 
\end{equation}
which comes from the $r$-components of the Maxwell equation.
We intend to use these dynamical equations \eqref{rrein}, \eqref{rmax}
along with one of the constraint equations in \eqref{seinconst} to solve for
the unknown functions $h_1(r)$, $k_1(r)$ and $w_1(r)$.

Solving \eqref{rrein} we get
\begin{equation}\label{hsol}
 h_1(r) = \frac{C_{h_1}^{1}}{r^4} + C_{h_1}^{2},
\end{equation}
where $C_{h_1}^{1}$ and $C_{h_1}^{2}$ are constants to be
determined. We can set $C_{h_1}^{2}$ to zero as it will lead to a
non-normalizable mode of the metric.  We then substitute the solution
for $h_1(r)$ from \eqref{hsol} into \eqref{rmax} and solve the
resultant equation for $w_1(r)$. The solution that we obtain is given
by
\begin{equation}\label{ksol}
 w_1(r) = C_{w_1}^1 r^2 + C_{w_1}^2 - q_0 \frac{C_{h_1}^{1}}{r^4}.
\end{equation}
Here again $C_{w_1}^1$, $C_{w_2}^2$ are constants to be determined. 
Again $C_{w_1}^1$ corresponds to a non-normalizable mode of the gauge field
and therefore can be set to zero.

Finally plugging in these solutions for $h_1(r)$ and $w_1(r)$ into one of the 
constraint equations in \eqref{seinconst} and then solving the 
subsequent equation we obtain
\begin{equation}
 k_1(r) = \frac{2}{3} r^3 \partial_i\beta^{(0)}_i + C_{k_1} 
              - \frac{2 q0}{r^2} C_{w_1}^2 + \left( \frac{2 {q_0}^2}{r^6}
               - \frac{m_0}{r^4} \right) C_{h_1}^1 
\end{equation}
Now the constants $C_{k_1}$ and $C_{w_1}^2$ may be absorbed into redefinitions 
of mass ($m_0$) and charge ($q_0$) respectively and hence may be set to zero. 
Further we can gauge away the constant $C_{h_1}^1$ by the following
redefinition of the  r coordinate
$$r \rightarrow r\left(1 + \frac{C}{r^4}\right),$$
$C$ being a suitably chosen constant.

Thus we conclude that all the arbitrary constants in this sector
can be set to zero and therefore our solutions may be summarized as
\be
h_1(r)=0,~~w_1(r)=0,~~k_1(r) = \frac{2}{3} r^3 \partial_i\beta^{(0)}_i\ .
\ee

In terms of the first order metric and gauge field this result reduces to
\be
\begin{split}
 \sum_{i} g^{(1)}_{ii}(r) &= 0, \\
g^{(1)}_{vv}(r) &= \frac{2}{3} r \partial_i\beta^{(0)}_i, \\
g^{(1)}_{vr}(r) &= 0,  \\
A^{(1)}_v(r) &= 0\ .
\end{split}
\ee
Now, we proceed to solving the equations in the vector sector.

\subsection{Vectors Of $SO(3)$ at first order}\label{vect1}

The vector components of metric and gauge field  $g^{(1)}$ 
and $A^{(1)}$ are parameterized by the functions $j_i^{(1)}(r)$ 
and $g_i^{(1)}(r)$ as follows
\begin{equation}
\label{vmetgagdef}
\begin{split}
g^{(1)}_{vi}(r) &= \left(\frac{m_0}{r^2}-\frac{q_0^2}{r^4}\right)
j_i^{(1)}(r)\\ 
A^{(1)}_{i}(r) &= -\left( \frac{\sqrt{3}q_0}{2
  r^2}\right) j_i^{(1)}(r) + g_i^{(1)}(r)
\end{split}
\end{equation}
Now we intend to solve for the functions $j_i^{(1)}(r)$ and $g_i^{(1)}(r)$.

\subsubsection*{\it \bf Constraint equations}\label{vecconst1}

The constraint equations in the vector sector comes only from the Einstein equation. 
So there is only one constraint equation in this sector. It is given by
\be \label{vecconsE}
g^{rr} E_{ri} + g^{rv} E_{vi}=0
\ee
which implies
\begin{equation}\label{veinconst}
 \partial_i m^{(0)} = - 4 m_0 \partial_v \beta_i^{(0)}.
\end{equation}
These equations also follow from the conservation of boundary stress
tensor at first order.  We shall use this constraint equation to
simplify the dynamical equations in the vector sector.

\subsubsection*{\it \bf Dynamical equations and their solutions}\label{sssec:vecdyn1}

In the vector sector we have two equations from Einstein equations 
(the $ri$ and $vi$-components) and one from Maxwell equations (the $i$th-component)
\footnote{Note that a linear combination of the $ri$ and $vi$-components of the 
Einstein equation appear as the constraint equation in \eqref{veinconst}.}.

The dynamical equation obtained from the $vi$-component of the Einstein equations
is given by
\begin{equation}
\label{einvec}
 \left(q_0^2-3m_0 r^2\right)\frac{d j_i^{(1)}(r)}{dr} + 4\sqrt{3} q_0
r^2 \frac{d g_i{(1)}(r)}{dr} + \left(m_0 r^2 - q_0^2\right) r \frac{d^2
j_i^{(1)}(r)}{dr^2} = -3 r^4 \partial_v \beta_i^{(0)}.
\end{equation}
Also the dynamical equation from the $i$th-component of the Maxwell equation
is given by
\begin{equation}
\label{maxvec}
\begin{split}
&r\left[2 \left(r^6-m_0
   r^2+q_0^2\right) \frac{d^2 g_i^{(1)}}{dr^2} r^2 + \left(6
   r^7+2 m_0 r^3-6 q_0^2 r\right)
   \frac{d g_i^{(1)}(r)}{dr}\right]\\
&-\sqrt{3} q_0 r\left(r^6-m_0
   r^2+q_0^2\right) \frac{d^2 j_i^{(1)}(r)}{dr^2}    
   +\sqrt{3}
   q_0 \left(r^6-3 m_0 r^2+ 5
   q_0^2\right) \frac{d j_i^{(1)}(r)}{dr}\\
&\qquad = \sqrt{3} ( q_0 \partial_v \beta_i^{(0)}
 + \partial_i q^{(0)}) r^3 - 24 q_0^2 \kappa r l^{(0)}_i,
\end{split}
\end{equation}
where $l_i$ is defined as
\be 
l_i \equiv \epsilon_{ijk}\partial_j \beta_k.
\ee

Now in order to solve this coupled set of differential equations
\eqref{einvec} and \eqref{maxvec} we shall substitute $g_i^{(1)}(r)$
obtained from \eqref{einvec} into \eqref{maxvec} and solve the
resultant equation for $j_i^{(1)}(r)$. For any function
$j_i^{(1)}(r)$, using \eqref{einvec} $g_i^{(1)}(r)$ may be expressed
as
\begin{equation}\label{gsol1}
 g_i^{(1)}(r) = (C_{g})_i+\frac{1}{4 \sqrt{3} q_0}\left(-\partial_v
   \beta_i^{(0)} r^3+4 m_0 j_i^{(1)}(r)-\frac{\left(m_0
   r^2-q_0^2\right) \frac{d j_i^{(1)}(r)}{dr}}{r}\right)\ .
\end{equation}
Here $(C_{g})_i$ is an arbitrary constant. It corresponds to non
normalizable mode of the gauge field and hence may be set to zero.

Substituting this expression for $g_i^{(1)}(r)$ into \eqref{maxvec} we
obtain the following differential equation for $j_i^{(1)}(r)$
\begin{equation}
 \begin{split}
 &\left(35 q_0^4+5 r^2 \left(r^4-6 m_0\right) q_0^2+3 m_0
r^4 \left(3 r^4+m_0\right)\right) \frac{d j_i^{(1)}(r)}{dr} \\
r &\left(-11 q_0^4-\left(5 r^6-14 m_0 r^2\right)
q_0^2-m_0 r^4 \left(r^4+3 m_0\right)\right) \frac{d^2 j_i^{(1)}(r)}{
dr^2}\\
& + r^2 \left(q_0^2-m_0 r^2\right) \left(r^6-m_0 r^2+q_0^2\right)
\frac{d^3j_i^{(1)}(r)}{dr^3}\\ & = \frac{1}{\sqrt{3}}\left(6 \sqrt{3} q_0
\partial_i q^{(0)} r^4+3 \sqrt{3} \partial_v \beta_i^{(0)} \left(5 r^6-m_0
r^2+q_0^2\right) r^4-144 \ r \ l_i^{(0)} q_0^3 \kappa \right)
\end{split}
\end{equation}
The solution to this equation is given by,
\begin{equation}
\begin{split}
 j_i^{(1)}(r)&=(C_{j}^1)_i + \frac{(C_{j}^2)_i
         r^2}{\frac{m_0}{r^2}-\frac{q_0^2}{r^4}} + \frac{r \ \partial_v
         \beta_i^{(0)}}{\frac{m_0}{r^2}-\frac{q_0^2}{r^4}}\\
         &\quad+\frac{ \sqrt{3} \ l_i^{(0)} \ 
         q_0^3 \kappa }{ m_0
         \left(\frac{m_0}{r^2}-\frac{q_0^2}{r^4}\right) r^4} + \frac{6
         r^2 q_0 (\partial_i q^{(0)}  + 3 q_0 \partial_v \beta_i^{(0)}
         )}{R^7
         \left(\frac{m_0}{r^2}-\frac{q_0^2}{r^4}\right)}
         F_1(\frac{r}{R},\frac{m_0}{R^4}),
\end{split}
\end{equation}
where again $(C_{j}^1)_i$ and $(C_{j}^2)_i$ are arbitrary
constants. $(C_{j}^2)_i$ corresponds to a non-normalizable mode of the
metric and so is set to zero. $(C_{j}^1)_i$ can be absorbed into a
redefinition of the velocities and hence is also set to zero.

Here the function $F_1(\frac{r}{R},\frac{m_0}{R^4})$ is given
by\footnote{Although the expression for
  $F_1(\frac{r}{R},\frac{m_0}{R^4})$ is very complicated but it
  satisfies some identities. One can use those identities to perform practical 
calculations with this function.}
\begin{equation}
 F_1(\rho, M) \equiv \frac{1}{3}\left( 1- \frac{M}{\rho^4} +
                \frac{Q^2}{\rho^6}\right) \int _{\rho}^{\infty} dp
                \frac{1}{\left( 1- \frac{M}{p^4} +
                \frac{Q^2}{p^6}\right)^2} \left(\frac{1}{p^8} -
                \frac{3}{4 p^7}\left( 1 + \frac{1}{M}\right) \right),
\end{equation}
where $Q^2 = M-1$.

Substituting this result for $j_i^{(1)}(r)$ into \eqref{gsol1} we
obtain the following expression for $g_i^{(1)}(r)$
\begin{equation}
\begin{split}
 &g_i^{(1)}(r) =
 \frac{\sqrt{3} r^3 \sqrt{R^2 \left(m_0-R^4\right)}}
 		{2 \left(m_0 (r-R) (r+R)+R^6\right)} 
                 (\partial_v \beta^{(0)}_i) 
                 + \frac{3 R^2 \kappa (m_0-R^4) 
                 }{2 \left(m_0
                  \left(r^2-R^2\right)+R^6\right)} l_i\\
                &-\frac{\sqrt{3} r^4 \left(r \left(m_0 \left(r^2-R^2\right)+R^6\right)
   F_1^{(1,0)}\left(\frac{r}{R},\frac{m_0}{R^4}\right)+\left(6 R^7-6 m_0 R^3\right)
   F_1\left(\frac{r}{R},\frac{m_0}{R^4}\right)\right)}{2 R^8 \left(m_0
   \left(r^2-R^2\right)+R^6\right)} (\partial_i q^{(0)} + 3 q_0 \partial_v \beta^{(0)}_i)
\end{split}
\end{equation}
where we use the notation $f^{(i,j)}(\alpha,\beta)$ to  denote the
partial derivative $ \p^{i+j}f/\p\alpha^i\p\beta^j$ of the function $f$.

Plugging back $j_i^{(1)}(r)$ and $g_i^{(1)}(r)$ back into
\eqref{vmetgagdef} we conclude that the first order metric and gauge
field in the vector sector is given by
\begin{equation}
\label{vmetgagres}
\begin{split}
g^{(1)}_{vi}(r) &= r \partial_v \beta_i^{(0)} +\frac{\sqrt{3} \ l_i^{(0)} q_0^3
  \kappa }{ m_0 r^4} + \frac{6 r^2}{R^7} q_0 (\partial_i
q^{(0)} + 3 q_0 \partial_v \beta_i^{(0)})
F_1(\frac{r}{R},\frac{m_0}{R^4})\\ 
A^{(1)}_{i}(r) &=
-\frac{\sqrt{3} r^5 F_1^{(1,0)}\left(\frac{r}{R},\frac{m_0}{R^4}\right)}{2 R^8}
(\partial_i q^{(0)} + 3 q_0 \partial_v \beta^{(0)}_i) +
\frac{3 R \kappa  \sqrt{m_0-R^4} \sqrt{R^2 \left(m_0-R^4\right)}}{2 m_0 r^2} l_i
\end{split}
\end{equation}
%

\subsection{Tensors Of $SO(3)$ at first order}\label{tens1}

The tensor components of the first order metric  is
parameterized by the function $\alpha_{ij}^{(1)}(r)$ such that
\begin{equation}
 g^{(1)}_{ij} = r^2 \alpha_{ij}^{(1)}.
\end{equation}
The gauge field does not have any tensor components therefore in this
sector there is only one unknown function to be determined.

There are no constraint equations in this sector and the only
dynamical equation is obtained from the $ij$-component of the Einstein
equation. This equation is given by
\begin{equation}
\label{teqn}
r \left(r^6-m_0 r^2+q_0^2\right) \frac{d^2 \alpha_{ij}(r)}{dr^2}
   -\left(-5 r^6+m_0 r^2+q_0^2\right) \frac{d \alpha_{ij}(r)}{dr} = -6
   \sigma_{ij}^{(0)} r^4
\end{equation}
where $\sigma_{ij}$ is given by
\begin{equation}
 \sigma_{ij}^{(0)} = \frac{1}{2}\left(\partial_i \beta_j^{(0)} +
         \partial_j \beta_i^{(0)} \right) - \frac{1}{3} \partial_k
         \beta_k^{(0)} \delta_{ij}.
\end{equation}

The solution to equation \eqref{teqn} obtained by demanding regularity
at the future event horizon and appropriate normalizability at
infinity. The solution is given by
\begin{equation}
 \alpha_{ij}^{(1)} = \frac{2}{R} \sigma_{ij}
 F_2(\frac{r}{R},\frac{m_0}{R^4}),
\end{equation}
where the function $F_2(\rho,M)$ is given by
\begin{equation}
 F_2(\rho,M) \equiv \int_{\rho}^{\infty}\frac{p \left(p^2+p+1\right)}{(p+1)
   \left(p^4+p^2-M+1\right)} dp
\end{equation}
with $ M\equiv m/R^4$ as before.

Thus the tensor part of the first order metric is determined to be
\begin{equation}
g^{(1)}_{ij} = \frac{2 r^2}{R} \sigma_{ij}
F_2(\frac{r}{R},\frac{m_0}{R^4}).
\end{equation}
%

\subsection{The global metric and the gauge field at first order}\label{fulmetgag1}

In this subsection, we gather the results of our previous sections to write 
down the entire metric and the gauge field accurate up to first order in the 
derivative expansion.

We obtain the metric as
\begin{equation}\label{firstorderAns:eq}
\begin{split}
ds^2 &= g_{AB} dx^A dx^B \\
&= -2 u_\mu dx^\mu dr - r^2\ V\ u_\mu u_\nu dx^\mu dx^\nu + r^2 P_{\mu\nu} dx^\mu dx^\nu \\
&\quad -2 u_\mu dx^\mu\ r \left[u^\lambda\partial_{\lambda}u_\nu-\frac{\partial_{\lambda}u^\lambda}{3}u_\nu\right]dx^\nu+ \frac{2 r^2}{R}F_2(\rho,M) \sigma_{\mu\nu} dx^\mu dx^\nu\\
&\quad -2 u_\mu dx^\mu\left[\frac{\sqrt{3}\kappa q^3}{m r^4}l_\nu+\frac{6qr^2}{R^7}P_\nu^\lambda\mathcal{D}_\lambda qF_1(\rho,M) \right] dx^\nu +\ldots\\
A &= \left[\frac{\sqrt{3} q}{2 r^2} u_\mu +\frac{3\kappa q^2}{2 m r^2}l_\mu -\frac{\sqrt{3}r^5}{2R^8}P_\mu^\lambda\mathcal{D}_\lambda q F_1^{(1,0)}(\rho,M) \right]dx^\mu +\ldots
\end{split}
\end{equation}
where $\mathcal{D}_{\lambda}$ is the weyl covarient derivative 
defined in appendix \ref{app:weylcov}. We also have defined
\begin{equation}
\begin{split}
& V\equiv 1-\frac{m}{r^4}+\frac{q^2}{r^6}\ ;\qquad l^\mu\equiv \epsilon^{\nu\lambda\sigma\mu}u_\nu\partial_\lambda u_\sigma ;\qquad P_\mu^\lambda\mathcal{D}_\lambda q \equiv P_\mu^\lambda\partial_\lambda q+ 3 (u^\lambda \partial_\lambda u_\mu) q ;\qquad \rho \equiv \frac{r}{R} \\
& \sigma^{\mu \nu} \equiv P^{\mu \alpha} P^{\nu \beta} \partial_{(\alpha} u_{\beta)}
-\frac{1}{3} P^{\mu \nu} \partial_{\alpha}u_{\alpha}
;\qquad M\equiv \frac{m}{R^4} ;\qquad Q\equiv \frac{q}{R^3} ;\qquad Q^2= M-1 \\
\end{split}
\end{equation}
and
\begin{equation}
\begin{split}
 F_1(\rho, M) &\equiv  \frac{1}{3}\left( 1- \frac{M}{\rho^4} + \frac{Q^2}{\rho^6}\right)
 \int _{\rho}^{\infty} dp \frac{1}{\left( 1- \frac{M}{p^4} + \frac{Q^2}{p^6}\right)^2}
 \left(\frac{1}{p^8} - \frac{3}{4 p^7}\left( 1 + \frac{1}{M}\right) \right)\\
F_2(\rho,M) &\equiv \int_{\rho}^{\infty}\frac{p \left(p^2+p+1\right)}{(p+1)
   \left(p^4+p^2-M+1\right)} dp \ .
\end{split}
\end{equation}
%

\subsection{The Stress Tensor and Charge Current at first order}\label{stresscur1}

In this section, we obtain the stress tensor and the charge current from 
the metric and the gauge field. The stress tensor can be obtained from 
the extrinsic curvature after subtraction of the appropriate counterterms 
\cite{Henningson:1998gx, Balasubramanian:1999re}.
We get the first order stress tensor as
\begin{equation}\label{stress:eq}
\begin{split}
T_{\mu\nu}= p(\eta_{\mu\nu}+4 u_\mu u_\nu) -2\eta \sigma_{\mu\nu} +\ldots
\end{split}
\end{equation}
where the fluid pressure $p$ and the viscosity $\eta$ are given by the expressions
\begin{equation}
\begin{split}
p \equiv \frac{M R^4}{16\pi G_5}\qquad;\qquad \eta \equiv \frac{R^3}{16\pi G_5}=\frac{s}{4\pi}
\end{split}
\end{equation}
where $s$ is the entropy density of the fluid obtained from the Bekenstein formula.

To obtain the charge current, we use
\begin{equation}
\begin{split}
J_{\mu} = \lim_{r\to\infty} \frac{r^2 A_\mu}{8\pi G_5} = n\ u_\mu -\mathfrak{D}\ P_{\mu}^\nu\mathcal{D}_\nu n + \xi\ l_\mu +\ldots
\end{split}
\end{equation}
where the charge density $n$, the  diffusion constant $\mathfrak{D}$ and 
an additional transport coefficient $\xi$ for the fluid under consideration are given by
\footnote{Here we have taken the chemical potential $\mu = 2 \sqrt{3} Q R$ which determines the 
normalization factor of the charge density $n$ (because thermodynamics tells us $n \mu = 4p -Ts$) 
which in turn determines the normalization of $J_\mu$. Note that due to the difference in $\mu$ with
\cite{Erdmenger:2008rm}, our normalization of $J_{\mu}$ is different from that in \cite{Erdmenger:2008rm}.}
\begin{equation}
\begin{split}
n\equiv \frac{\sqrt{3} q}{16\pi G_5} \quad;\quad \mathfrak{D}=\frac{1+M}{4MR} \quad;\quad   \xi\equiv \frac{3 \kappa q^2}{16\pi G_5 m}
\end{split}
\end{equation}

We note that when the bulk Chern-Simons coupling $\kappa$ is non-zero,
apart from the conventional diffusive transport, there is an additional
non-dissipative contribution to the charge current which is proportional
to the vorticity of the fluid. To the extent we know of, this is a hitherto
unknown effect in the hydrodynamics which is exhibited by the conformal
fluid made of $\mathcal{N}=4$ SYM matter. It would be interesting to find
a direct boundary reasoning that would lead to the presence of such a
term - however, as of yet, we do not have such an explanation and we hope
to return to this issue in future.

The presence of such an effect was indirectly observed by the
authors of \cite{Bhattacharyya:2007vs} where they noted a discrepancy
between the thermodynamics of charged rotating AdS black holes and the
fluid dynamical prediction with the third term in the charge current absent.
We have verified that this discrepancy is resolved once we take into
account the effect of the third term in the thermodynamics of the
rotating $\mathcal{N}=4$ SYM fluid. In fact, one could go further
and compare the first order metric that we have obtained with
rotating black hole metrics written in an appropriate gauge. We
have done this comparison up to first order and we find that
the metrics agree up to that order.

\section{Second Order Hydrodynamics} \label{secord}

In this section we will find out the metric, stress tensor and charge
current at second order in derivative expansion. We will follow the
same procedure as in \cite{Bhattacharyya:2008jc} but in presence charge parameter
$q$. 

The metric and gauge field perturbations at second order that we consider are 
\be \label{metsecper}
g^{(2)}_{\alpha \beta}dx^{\alpha}dx^{\beta} 
= -3 h_2(r) dv dr + r^2 h_2(r) dx^i dx_i + \frac{k_2(r)}{
r^2} dv^2 + 12 r^2 j_i^{(2)}dv dx^i + r^2 \alpha_{ij}^{(2)}dx^i dx^j 
\ee
and
\ben\label{gaugsecper}
A^{(2)}_v &=& -\frac{\sqrt{3}}{ 2r^2}  w_2(r) \nn
A^{(2)}_i &=& \frac{\sqrt{3}}{2}
r^5 g_i^{(2)}(r) dx^i\ .
\een
Here we have used a little different parameterizations (from first
order) for  metric and gauge field perturbations in the vector
sector. We found that this aids in writting the corresponding dynamical 
equations for $j_i^{(2)}(r)$ and  $g_i^{(2)}(r)$  in a more tractable
form (as we will see later).

Like neutral black brane case, here also we will list all the source
terms (second order in derivative expansion) which will appear on the
right hand side of the constraint dynamical equations in scalar, vector
and tensor sectors. These source terms are built out of second
derivatives of $m$, $q$ and $\beta$ or square of first derivatives of
these three fields. We can group these source terms according to their
transformation properties under ${\bf SO(3)}$ group. A complete list
has been provided in table \ref{tablevtdefs}. In the table the quantities
$l_i$ and $\sigma_{ij}$ are defined to be
\be
l_i= \epsilon_{ijk}\p_j \beta_k ~, ~~~~~~ \sigma_{ij} = \frac{1}{2}
(\p_i \beta_j + \p_j \beta_k) -\frac{1}{3} \delta_{ij} \p_k \beta_k\ .
\ee
%

%
\TABLE{
\caption{An exhaustive list of two derivative terms in made up from
the mass, charge and velocity fields. In order to present the results
economically, we have dropped the superscript on the velocities
$\beta_i$ charge $q$ and the mass $m$, leaving it implicit that
these expressions are only valid at second order in the derivative
expansion.}
\label{tablevtdefs}
 \begin{tabular}{|l|l|l|}
\hline & & \\ \;\;\; {\bf 1} of $SO(3)$ &\;\;\;\;\; {\bf 3} of $SO(3)$
&\;\;\; \;\;\;\;\;{\bf 5} of $SO(3)$ \\ & &\\ \hline \hline & &\\
\;\;\;S1 = $\partial_v^2 m$ & \;\;\;$\text{V1}_i =
\partial_i\partial_v m$ & \;\;\;$\text{T1}_{ij} = \partial_i\partial_j
m -\frac{1}{3}$ s3 $\delta_{ij}$ \\ & &\\ \;\;\;S2 =
$\partial_v\partial_i\beta_i$ & \;\;\;$\text{V2}_i =
\partial_v^2\beta_i$ & \;\;\;$\text{T2}_{ij} = \partial_{(i} l_{j)}$
\\ & &\\ \;\;\;S3 = $\partial^2 m$ & \;\;\;$\text{V3}_i = \partial_v
l_i $ & \;\;\;$\text{T3}_{ij} = \partial_v \sigma_{ij}$ \\ & &\\
\;\;\;ST1 = $\partial_v\beta_i\,\partial_v\beta_i$ &
\;\;\;$\text{V4}_i = \frac{9}{5}\partial_j \sigma_{ji} -
\partial^2\beta_i$ & \;\;\;$\text{TT1}_{ij} =
\partial_v\beta_i\,\partial_v\beta_j - \frac{1}{3}$ ST1 $\delta_{ij}$
\\ & &\\ \;\;\;ST2 = $l_i\,\partial_v\beta_i$ & \;\;\;$\text{V5}_i =
\partial^2\beta_i$ & \;\;\;$\text{TT2}_{ij} = l_{(i} \,
\partial_v\beta_{j)} - \frac{1}{3}$ ST2 $\delta_{ij} $ \\ & &\\
\;\;\;ST{3} = $\left(\partial_i\beta_i\right)^2$ & \;\;\;$\text{VT1}_i
= \frac{1}{3}(\partial_v\beta_i)(\partial_j\beta^j)$ & \;\;\;$
\text{TT3}_{ij} =2\, \epsilon_{kl(i} \, \partial_v\beta^k\,
\partial_{j)} \beta^l+ \frac{2}{3}$ ST2 $\delta_{ij}$ \\ & &\\
\;\;\;$\text{ST4} = l_i\,l^i $ & \;\;\;$\text{VT2}_i =
-\epsilon_{ijk}\, l^j\, \partial_v\beta^k $ & \;\;\;$\text{TT4}_{ij}
= \partial_k\beta^k\, \sigma_{ij} $ \\ & &\\ \;\;\;$\text{ST5} =
\sigma_{ij}\,\sigma^{ij}$ & \;\;\;$\text{VT3}_i =
\sigma_{ij}\,\partial_v\beta^j$ & \;\;\;$\text{TT5}_{ij} = l_i \,l_j
- \frac{1}{3}$ ST4 $\delta_{ij}$ \\ & &\\ \;\;\;$\text{QS1} =
\partial_v^2 q$ & \;\;\;$\text{VT4}_i = l_i\, \partial_j\beta^j$ &
\;\;\;$\text{TT6}_{ij} = \sigma_{ik}\, \sigma^k_j - \frac{1}{3}$ ST5
$\delta_{ij}$ \\ \;\;\;& &\\ \;\;\;$\text{QS2} = \partial_i \partial_i
q$& \;\;\;$\text{VT5}_i = \sigma_{ij}\,l^j$ & \;\;\;$ \text{TT7}_{ij}
= 2\, \epsilon_{mn(i} \, l^m \,\sigma^n_{j)} $\\ & &\\
\;\;\;$\text{QS3} = l_i \partial_i q$ & \;\;\;$\text{QV1}_i =
\partial_i \partial_v q$ & \;\;\;$\text{QT1}_{ij} =
\partial_i\partial_j q -\frac{1}{3}$ QS2 $\delta_{ij}$ \\ & &\\
\;\;\;$\text{QS4}=(\partial_i q)^2$ & \;\;\;$\text{QV2}_i = \partial_i
q \partial_k \beta^k$ & \;\;\;$\text{QT2}_{ij} = \partial_{(i}q l_{j)}
- \frac{1}{3} $ QS3 $\delta_{ij}$\\ & &\\ \;\;\;$\text{QS5}=
(\partial_i q) (\partial_v \beta_i$) & \;\;\;$\text{QV3}_i = \epsilon_{ijk}
\partial_j l_k$ & \;\;\;$\text{QT3}_{ij} = \partial_{(i}q
\partial_{j)}q - \frac{1}{3} $ QS4 $\delta_{ij}$\\ & &\\ &
\;\;\;$\text{QV4}_i = \sigma_{ij} \partial_j q$ &
\;\;\;$\text{QT4}_{ij} = \partial_{(i}q \partial_{v}\beta_{j)} -
\frac{1}{3}$ QS5 $\delta_{ij}$\\ & &\\ & \;\;\;$\text{QV5}_i =
\epsilon_{ijk} \partial_v \beta_j \partial_k q$ 
&\;\;\;$\text{QT5}_{ij}= \epsilon_{(ikm}\partial_k q\ \sigma_{mj)}$\\ & &\\ \hline
\end{tabular} 
}
%

In table \ref{tablevtdefs} we have already employed the first order
conservation relations i.e. equation \ref{1stordcon1} and \ref{1stordcon2}.
Using these two relations we have eliminated the first derivatives of
$m$ and $q$. However at second order in derivative expansion we also have the relations
\be \label{2stordcon1}
\p_{\mu} \p_{\nu}T^{\mn}_{(0)}=0 \ ,
\ee
and
\be 
\label{2stordcon2} 
\p_{\lambda}\p_{\mu} J^{\mu}_{(0)}=0 \ .
\ee 

The equations \eqref{2stordcon1} and \eqref{2stordcon2} imply some relations between 
the second order source terms which are listed in table \ref{tablevtdefs}. These 
relations are
\ben 
\text{S1} &=& \frac{\text{S3}}{3} - \frac{8 m}{3} \text{ST1} +
\frac{16 m}{9} \text{ST3} - \frac{2 m }{3} \text{ST4} + \frac{ 4 m
}{3} \text{ST5}\nn \text{S2}&=& - \frac{1}{4 m} \text{S3} + 4
\text{ST1} + \frac{1}{2} \text{ST4} - \text{ST5}\nn \text{QS1} &=&
q\left(- \text{ST1} - \text{S2} + \text{ST3}\right ) - \text{QS5}\nn
\text{V1}_i &=& m \left( -\frac{40}{9} \text{V4}_i -\frac{4 }{ 9}
\text{V5}_i + \frac{56 }{3} \text{VT1}_i + \frac{4 }{ 3} \text{VT2}_i +
\frac{8}{3} \text{VT3}_i \right) \nn \vtwo_i &=& \frac{10}{9} \vfour_i + \frac{1}{9}
\vfive_i - \frac{2}{3} \vtone_i + \frac{1}{6} \vttwo_i - \frac{5}{3} \vtthre_i
\nn 
\vthre_i &=& -\frac{1}{3} \vtfour_i + \vtfive_i \nn \qvone_i &=& -
q \left (\frac{10}{3} \vfour_i + \frac{1}{2} (\vttwo_i + 2 \vtone_i + 2
\vtthre_i) + \frac{1}{3} \vfive_i \right ) \nn && - \qvtwo_i - \frac{1}{2}
\left ( 2 \qvfour_i + \qvthre_i + \frac{2}{3} \qvtwo_i \right )\nn \text{T1}_{ij}
&=& -4 m\left(\text{T3}_{ij}+\frac{1}{4}
\text{TT5}_{ij}-4\text{TT1}_{ij}+\frac{1}{3}\text{TT4}_{ij}+\text{TT6}_{ij} \right) 
\een
With these relation between the source terms we will now solve the
Einstein equations and Maxwell equations to find out the constraint
and dynamical equations at second order in derivative expansion. As
in the first order calculations we shall perform this seperately
in various sectors denoting different representation of the 
boundary rotation group $SO(3)$.

\subsection{Scalars of $SO(3)$ at second order}

We parametrise the metric and the gauge field as follows
\begin{equation}
\begin{split}
\sum_{i} g^{(2)}_{ii}(r) &=  3 r^2 h_2(r), \\
g^{(2)}_{vv}(r) &= \frac{k_2(r)}{r^2} \\
g^{(2)}_{vr}(r) &= -\frac{3}{2} h_2(r) \\
A^{(2)}_v(r) &= -\frac{\sqrt{3} w_2(r)}{2 r^2}.
\end{split}
\end{equation}
Now we intend to solve for the functions $h_2(r),k_2(r)$ and $w_2(r)$.

\subsubsection*{\it \bf Constraint Equations}\label{scalconst2}
%
As we have already explained, there are three constraint
equations. First two come from Einstein equations (Eq. \ref{sclconsE1}
and \ref{sclconsE1}) and the third one comes from Maxwell equations
(Eq. \ref{sclconsM1}). The first constrain from Einstein equations
gives
\be
\p_vm^{(1)} = \frac{2}{3} R^3 \ \stfive 
\ee
Second constraint implies relation between $k_2(r)$
and $h_2(r)$. This constraint equation is given by
\be \label{consteq}
-m_0 h_2'(r)+3 r^4 h_2'(r)+12 r^3
   h_2(r)-k_2'(r)+\frac{4 q_0
   w_2(r)}{r^3}-\frac{2 q_0 w_2'(r)}{r^2} = S_C,
\ee 
where the source term $S_C$ is given in appendix \ref{appscl2}.

The constraint relation coming from Maxwell equations is given by
\ben
\p_vq^{(1)} &=& -\frac{3 \text{$q_0$} \left(\text{$R$}^4+\text{$m_0$}\right)
  }
{16 \text{$m_0$}^2 \text{$R$}} \text{S3}
+\frac{ \left(\text{$R$}^4+\text{$m_0$}\right)}{4
   \text{$m_0$} \text{$R$}}\text{QS2}-\frac{6 \sqrt{3} \text{$q_0$}^2 
 \kappa
   }{\text{$m_0$}} \text{ST2} \nn
&-& \frac{  \left(\text{$m_0$}-11
   \text{$R$}^4\right)}{4 \text{$m_0$} \text{$R$}}\text{QS5}
-\frac{2 \sqrt{3} \text{$q_0$}  \kappa
   }{\text{$m_0$}}\text{QS3} -\frac{\text{$q_0$} }{4 \text{$m_0$} 
\text{$R$}^3}\text{QS4} \nn
&+&\frac{9 \text{$q_0$} \left(3 \text{$R$}^4+\text{$m_0$}\right)}{4
   \text{$m_0$} \text{$R$}} \stone
\een
%

\subsubsection*{\it \bf Dynamical Equations and their solutions}\label{scaldyn2}

The Dynamical Equations in the scalar sector (coming from the 
Einstein equation $E_{rr}=0$) is given by 
\be \label{eqh}
r h_2''(r) + 5 h_2'(r) = S_h \ .
\ee
 The source term $S_h$ is explicitly given in appendix \ref{appscl2}.

The second dynamical scalar equation, which comes form the Maxwell equations ($M(r)=0$),
is given by
%
\ben \label{eqw}
-6 q_0 h_2'(r) + r w_2''(r)- w_2'(r) = S_M(r).
\een
The explicit form of the source term $S_M(r)$ is again given in appendix \ref{appscl2}.

The source terms have the same large $r$ behavior as uncharged case (see \cite{Bhattacharyya:2008jc})
because the charge dependent terms (leading) are more suppressed than
that of charge independent terms. So one can follow the same procedure
to obtain the solution for $h_2(r)$ and $k_2(r)$. Here we present the 
result schematically. Firstly, we solve equation \eqref{eqh} for the 
function $h_2(r)$; we obtain
\begin{equation}
 h_2(r) = \int \left( \frac{1}{r^5} \left( \int \left( r^4 S_h(r)\right) dr 
           + C_h^{(1)} \right) \right) dr + C_h^{(2)},
\end{equation}
where $C_h^{(1)}$ and $C_h^{(2)}$ are the constants of integration.
We then plug in this solution for $h_2(r)$ in to \eqref{eqw}. Solving
the resultant equation for the $w_2$ we obtain,
\begin{equation}
 w_2(r) = \int \left( r \left( \int \left( \frac{1}{r^2} S_w(r) \right) dr +C_w^{(1)}
           \right) \right) dr + C_w^{(2)},
\end{equation}
where again  $C_w^{(1)}$ and $C_w^{(2)}$ are integration constants, and
the function $S_w(r)$ is 
$$ S_w(r) = S_M(r) + 6 q_0 h_2'(r).$$
Finally, we substitute the functions $h_2(r)$ and $w_2(r)$ solved above,
in to \eqref{consteq} to obtain the following equation for $k_2(r)$
\begin{equation}
 k_2'(r) = (3 r^4 - m_0) h_2'(r) + 12 r^3 h_2(r) + \frac{4 q_0}{r^3} w_2(r) 
           - \frac{2 q_0}{r^2} w_2'(r) - S_C \equiv S_k(r).
\end{equation}
This equation can be easily integrated to obtain 
\begin{equation}
 k_2(r) = \int S_k(r) dr + C_k,
\end{equation}
$C_k$ being the integration constant.
All the integration constants in the above solutions are 
obtained by imposing regularity at the horizon and normalizability of the
functions, just as in the first order computation.
 
\subsection{Vectors of $SO(3)$ at second order}\label{vect2}

As given in \eqref{metsecper} and \eqref{gaugsecper}, in this sector we parametrize\footnote{Note that
the parametrization of the gauge field at this order is different 
from the one used for the scalar sector.}
the metric, and the gauge field respectively in the following way
\begin{equation}
\begin{split}
 g_{vi} &= 6 r^2 j^{(2)}_i(r)\\
 A_i^{(2)} &= \frac{\sqrt{3}}{2} r^2 g^{(2)}_i (r).
\end{split}
\end{equation}

\subsubsection*{\it \bf Constraint Equations}\label{vecconst2}

In this sector, the constraint equation comes only from the Einstein equations (\ref{vecconsE}).
This constraint relation is give by
\ben
\p_im^{(1)} &=& \frac{10  \text{$R$}^3}{9}\text{V4}_i +\frac{10 
   \text{$R$}^3}{9}\text{V5}_i + \frac{10 
   \text{$R$}^3}{3}\text{VT1}_i - \frac{5 
   \text{$R$}^3}{6}\text{VT2}_i \nn
&+&\frac{6 \text{$q_0$} 
   \text{$R$}}{\text{$m_0$} - 3 \text{$R$}^4}\text{QV4}_i - \frac{\left(21
   \text{$R$}^7-43 \text{$m_0$} \text{$R$}^3\right)
   }{3 \left(\text{$m_0$}-3 \text{$R$}^4\right)}\text{VT3}_i\ .
\een
%

\subsubsection*{\it \bf Dynamical Equations and their solutions}\label{sssec:vecdyn2}

There are two vector dynamical equations. The first equation 
comes from Einstein equation and is given by
\begin{equation}\label{vecdyn1}
q_0 r {g^{(2)}_{i}}'(r) + 5 q_0 {g^{(2)}_{i}}(r) + r {j^{(2)}_{i}}''(r)+5
  {j^{(2)}_{i}}'(r) = (S_E^{\text{vec}})_i(r),
\end{equation}
where $(S_E^{\text{vec}})_i(r)$ is the source terms given in the appendix
\ref{appvec2}. The second dynamical equation comes from Maxwell
equation and is given by
\begin{equation}\label{vecdyn2}
\begin{split}
 \sqrt{3} \Big(-m_0 r^4 {g^{(2)}_{i}}''(r) &+ q_0^2 r^2 {g^{(2)}_{i}}''(r) + r^8
   {g^{(2)}_{i}}''(r)+ {g^{(2)}_{i}}'(r) \left(-9 m_0 r^3+7 q_0^2 r+13
   r^7\right)    \\ &  +5  {g^{(2)}_{i}}(r) \left(-3 m_0 r^2+q_0^2+7
   r^6\right)+12 q_0 {j^{(2)}_{i}}'(r)\Big) = (S_M^{\text{vec}})_i(r)
\end{split}
\end{equation}
where  $(S_M^{\text{vec}})_i(r)$ is the other source term the explicit form of which is also
given in the appendix \ref{appvec2}.
The sources $(S_M^{\text{vec}})_i(r)$ and $(S_E^{\text{vec}})_i(r)$ are expressed in terms
of the weyl invariant quantities $(W_v)_i^m$ which are defined in appendix \ref{app:weylcov}.
We can now solve equation \eqref{vecdyn1} for the function $g^{(2)}_i(r)$ to obtain
\begin{equation}\label{gsol2}
 g^{(2)}_i(r) =  -\frac{{j^{(2)}_i}'(r)}{q_0} + \frac{(W_v)_i^1+(W_v)_i^2}{6 q_0 r^3} 
                 -\left( \frac{1}{q_0 r^5}\right)\int_r^{\infty} x^4 \left((S_E^{\text{vec}})_i(r) 
                 - \frac{(W_v)_i^1+(W_v)_i^2}{3 x^3} \right) dx,
\end{equation}
where the integrating constant has been chosen by the normalizability condition.
Plugging in this solution in to \eqref{vecdyn2} we obtain the following
effective equation for $j^{(2)}_i(r)$
\begin{equation}\labell{eqj2}
 \frac{d}{dr} \left( \frac{1}{r} \frac{d}{dr} \left( r^7 
 \left( V^{(0)}(r) \right)^2 \frac{d}{dr} 
 \left( \frac{1}{V^{(0)}(r)} j^{(2)}_i(r) \right) \right) \right) + S_i(r)=0,
\end{equation}
where
\begin{equation}
\begin{split}
 S_i(r) & =\left( -\frac{1}{\sqrt{3} r^2} \right) \Bigg( \sqrt{3} \left(r \left(m_0
   \left(R^2-r^2\right)+r^6-R^6\right)
   (S_E^{\text{vec}})_i'(r) \right. \\ & \quad \quad  \left. + (S_E^{\text{vec}})_i(r) \left(m_0
   \left(R^2-3 r^2\right)+7
   r^6-R^6\right)\right)-\sqrt{R^2
   \left(m_0-R^4\right)}
   (S_M^{\text{vec}})_i(r) \Bigg).
\end{split}
\end{equation}

Finally, the solution to the equation \eqref{eqj2} is given by
\begin{equation}\label{jsol2}
\begin{split}
 j^{(2)}_i(r) &= - V^{(0)}(r)\int_r^{\infty} \frac{1}{x^7 \left(V^{(0)}(x)\right)^2} \Bigg( 
        \int_x^{\infty} y \int_y^{\infty} S^{reg}_i(z) dz dy \Bigg) dx
            \\ & -V^{(0)}(r)\int_r^{\infty} \frac{1}{x^7 \left(V^{(0)}(x)\right)^2} \Bigg[ C_i^{(j)} 
                - \frac{1}{3(m_0 3 R^2)}3 R^7 \Bigg( \left( (W_v)_i^1 + (W_v)_i^4 \right) x
                 \\ &  - m_0 R^3 \left((W_v)_i^1 + 3(W_v)_i^4 \right) x
                -\frac{1}{2} m_0 \left( (W_v)_i^1 +(W_v)_i^2\right) x^4
                +\frac{3}{2} R^4 \left( (W_v)_i^1 +(W_v)_i^2 \right) x^4 \Bigg)\Bigg] dx,
\end{split}
\end{equation}
where again for convenience we have defined
\begin{equation}
 S^{reg}_i(z) = \frac{R^3 \left(m_0 ((W_v)_i^1+3 (W_v)_i^4)-3
   R^4 ((W_v)_i^1+(W_v)_i^4)\right)}{3 z^2
   \left(m_0-3
   R^4\right)}-S_i(z)-\frac{4}{3} z
   ((W_v)_i^1+(W_v)_i^2).
\end{equation}
The constant  $C_i^{(j)}$ is determined by the regularity at horizon and is given by
\begin{equation}\label{Csol}
\begin{split}
 C_i^{(j)} &= -\frac{1}{12
   m_0 \left(m_0-3 R^4\right)} \Bigg( R^4 \Big(m_0^2 (9 (W_v)_i^1+4 (W_v)_i^2+15
   (W_v)_i^4) \\ &  -6 m_0 R^4 (6 (W_v)_i^1+3 (W_v)_i^2+4
   (W_v)_i^4)+9 R^8 (3 (W_v)_i^1+2
  (W_v)_i^2+(W_v)_i^4)\Big) \\ & -9 R^2 \left(m_0^2-4
   m_0 R^4+3 R^8\right)
   \left(\int_{R}^{\infty } S^{reg}_i(x) \, dx\right)+6
   m_0 \left(m_0-3 R^4\right)
   \int_{R}^{\infty } y^2 S^{reg}_i(y) \, dy \Bigg),
\end{split}
\end{equation}

We now have to plug in the source terms (given in 
Appendix \ref{appvec2}) and perform the integrals to 
write the solutions explicitly. Since such explicit solution
would be very complicated, we do not provide it here. Nevertheless,
from the above solution we extract the boundary charge current as
we explicate in the following section.

%
\subsection{Boundary Charge Current at second order}\label{chcur2}
%
The charge current at second order in derivative expansion is given by
\be
J_{\mu}^{(2)} =\lim_{r \to \infty} \frac{r^2 A_{\mu}^{(2)}}{8\pi G_5}\ .
\ee
The gauge field perturbation at this order is parametrised by the function
$g^{(2)}_i(r)$. Thus to obtain the charge current density we have to consider the
asymptotic limit (i.e. the $r\rightarrow \infty$ limit) of the
function $g^{(2)}_i(r)$. This function is given by \eqref{gsol2}. The function
$j^{(2)}_i(r)$ in that equation is in turn given by \eqref{jsol2}.

If we carefully extract the coefficient of the $1/r^2$ term in the 
$r \rightarrow \infty$ limit of the gauge field (using the equation referred to 
in the last paragraph) we find that the charge current is given by
\begin{equation}
 J_{i}^{(2)} = \frac{m_0 (W_v)^2_i-6 C^{(j)}_i}{4 \sqrt{3}
   \sqrt{R^2 \left(m_0-R^4\right)}},
\end{equation}
the constant $C^{(j)}_i$ being given by the equation \eqref{Csol}.
Plugging in the sources in to equation \eqref{Csol} and performing the
integrations we find

\begin{equation}
 J_{i}^{(2)} =\left( \frac{1}{8 \pi G_5}\right)  \sum_{l=1}^5 \mathcal{C}_l (W_v)^l_i,
\end{equation}
where the coefficients of the Weyl invariant terms $(W_v)^l_i$
are given by \footnote{All these coefficients perfectly match with the corresponding coefficients
in version 4 of \cite{Erdmenger:2008rm} }
\begin{equation}
 \begin{split}
  \mathcal{C}_1 & = \frac{3 \sqrt{3} R \sqrt{M-1}}{8 M}, \\
  \mathcal{C}_2 & =  \frac{\sqrt{3} R (M-1)^{3/2} }{4 M^2},\\
  \mathcal{C}_3 & = - \frac{3 R  \kappa (M-1)  }{2 M^2},\\
  \mathcal{C}_4 & = \frac{1}{4} \sqrt{3} R  \sqrt{M-1}  \log (2) 
                   + \mathcal{O}(M-1),\\
  \mathcal{C}_5 & = -\frac{\sqrt{3}  R \sqrt{M-1} 
                    \left(M^2-48 (M-1) \kappa ^2+3\right)}{16 M^2}. \\
 \end{split}
\end{equation}
We have expressed the above results in terms of the parameters $M$ and $R$
with $M=m_0/R^4$.

\subsection{Tensors Of $SO(3)$ at second order}\label{tens2}

We now consider the tensor modes at second order. Following the first order calculations we pametrize the 
traceless symmetric tensor components of the second order metric by the function $\alpha^{(2)}_{ij}(r)$ 
such that
\begin{equation}
g^{(2)}_{ij} = r^2 \alpha^{(2)}_{ij}(r).
\end{equation}
In this sector there are no constraint equations. However, there is a dynamical equation which we 
solve in the following subsection. 


\subsubsection*{\it \bf Dynamical equations and their solutions}\labell{tensordyneq}

The $ij$-component of the Einstein equation gives the dynamical equation 
for $\alpha^{(2)}_{ij}(r)$ which is similar to \eqref{teqn}. However 
the source term of the differential equation is modified in the second order.
Thus, at second order this equation is given by
\begin{equation}
\label{dyneqten}
-\frac{1}{2 r} \frac{d}{d r} \left( \frac{1}{r} \left( q_0^2
-m_0 r^2 +r^6 \right) \frac{d}{d r} \alpha^{(2)}_{ij}(r) \right) = {\bf T}_{ij}(r),
\end{equation}
where we write the source in terms of weyl-covariant quantities as follows 
\begin{equation}
\begin{split}
{\bf T}_{ij}(r) = \sum_{l=1}^9 \tau_{l}(r) ~WT^{(l)}_{ij}.
\end{split}
\end{equation}
We define the weyl-covariant terms $WT^{(l)}_{ij}$ in appendix \ref{app:weylcov}.
The coefficients $\tau_l(r)$ of these weyl-covariant terms are given in appendix \ref{appten2}.

The solution to \eqref{dyneqten} which is regular at the outer horizon 
and normalizable at infinity is given by
\begin{equation}
\label{alpha2sol}
\alpha^{(2)}_{ij}(r) =  \int_{r}^{\infty} \left( \left(\frac{\xi}{q_0^2
           -m_0 \xi^2 + \xi^6}\right)  \int_1^{\xi} \left(
           ~2 ~\zeta ~{\bf T}_{ij}(\zeta) \right) d\zeta \right) d\xi.
\end{equation}
We need to plug in the source from appendix \ref{appten2} in to the
above equation and perform the integrals to obtain an explicit answer.
However, as in the second order vector sector this turns out to be
very complicated in general and therefore we do not produce it here.
The transport coefficients, however, of the boundary stress tensor  
 at second order in derivative expansion may be
obtained only by knowing the function $\alpha^{(2)}_{ij}(r)$
asymptotically (near the boundary). In the next subsection,
we compute this boundary stress tensor.

\subsection{Boundary Stress Tensor at second order} \label{Stressten2}

As mentioned earlier in subsection \ref{stresscur1},
the AdS/CFT prescription for obtaining the boundary 
stress tensor from the bulk metric is given by
\begin{equation}\label{bdystten}
T^{\mu}_{\nu} = -\frac{1}{8 \pi G_5} \lim_{r \rightarrow \infty} \Big( r^4 
                \left( K^{\mu}_{\nu} - \delta^{\mu}_{\nu}\right) \Big),
\end{equation}
where $K^{\mu}_{\nu}$ is the extrinsic curvature normal to the constant r surface.
Now, as is apparent from the formula, we need to know the asymptotic expansion
of the metric perturbation $\alpha^{(2)}_{ij}(\rho)$ in order to obtain the
stress tensor. The asymptotic expansion of the solution \eqref{alpha2sol} for 
$\alpha^{(2)}_{ij}(\rho)$ is given by
\begin{equation}
 \alpha^{(2)}_{ij}(\rho) = \frac{1}{r^2} \left( WT^{(3)}_{ij} - \frac{1}{2} WT^{(2)}_{ij} 
                           -\frac{1}{4} WT^{(4)}_{ij}\right)
                           + \frac{1}{4 r^4} \sum_{l=1}^9 \mathcal{N}_{l} ~WT^{(l)}_{ij} 
                           + \mathcal{O}\left(\frac{1}{r^5} \right),
\end{equation}
The leading term of this asymptotic expansion gives divergent contributions 
to the stress tensor which are canceled by divergence arising from the 
expansion of $g^{(0)} + g^{(1)}$ up to second order. 

On plugging in this asymptotic solution for the metric in to the formula 
\eqref{bdystten} we obtain
\begin{equation}
 T_{\mu \nu} = \left( \frac{1}{16 \pi G_5}\right)\sum_{l=1}^9 \mathcal{N}_{l} ~WT^{(l)}_{\mu \nu}.
\end{equation}
with  $\mathcal{N}_{l}$ being the transport coefficients
at second order in derivative expansion. These transport coefficients are given by 
%
\begin{equation}
\begin{split}
\mathcal{N}_{1} &= R^2 \left(\frac{M}{\sqrt{4 M-3}}  
          \log \left(\frac{3-\sqrt{4 M-3}}{3+\sqrt{4 M-3}}\right)+2\right) \Bigg),\\
\mathcal{N}_{2} &= -\frac{M R^2}{2 \sqrt{4 M-3}}\log \left(\frac{3-\sqrt{4 M-3}}{\sqrt{4 M-3}+3}\right),\\
\mathcal{N}_{3} &= 2 R^2,\\
\mathcal{N}_{4} &= \frac{R^2}{M} (M-1) \left(12 (M-1) \kappa ^2-M\right),\\
\mathcal{N}_{5} &= -\frac{(M-1) R^2}{2 M},\\
\mathcal{N}_{6} &= \frac{1}{2} (M-1) R^2 \Big(\log (8)-1\Big) + \mathcal{O}\left((M-1)^2\right),\\
\mathcal{N}_{7} &= \frac{\sqrt{3} (M-1)^{3/2} R^2 \kappa }{M},\\
\mathcal{N}_{8} &= 0\\
\mathcal{N}_{9} &= 0.
\end{split}
\end{equation}

\section{Discussion}

In this paper, we have computed the metric dual to a fluid with a globally 
conserved charge and used that to find the energy-momentum  tensor and the 
charge current in arbitrary fluid configurations to second order in the 
boundary derivative expansion.  

Note that the corresponding construction of the bulk metric dual to an 
uncharged fluid flow was characterized by a great deal of universality. 
This universality had its origin in the fact that every two derivative 
theory of gravity (interacting with other fields) that admits $AdS_5$ as 
a solution, also admits a consistent truncation to the equations of Einstein
gravity with a negative cosmological constant. This universality 
does not extend to Einstein Maxwell system. The possibility of extending 
the Einstein Maxwell system by a Chern Simon's term of arbitrary coefficient, 
accounted for in this paper, is an illustration of the reduced universality
of our calculations. 

We have seen that a nonzero value for the coefficient of the 
Chern-Simons in the bulk leads to an interesting dual hydrodynamic effect 
(recall that this coefficient is indeed nonzero in strongly 
coupled ${\cal N}=4$ Yang Mills ). At first order in the derivative 
expansion we find that the charge current has a term proportional to 
$l^\alpha\equiv\epsilon^{\mu\nu\lambda\alpha}u_\mu\partial_\nu u_\lambda$ 
in addition to the more familiar Fick type diffusive term. It would be 
interesting to reproduce this term from a computation directly in the 
boundary. In more general terms, it would be interesting to gain better 
intuition for effect induced by the bulk Chern-Simons term on boundary 
dynamics.

We have already remarked on the fact that the pseudo-tensor terms
appearing in the stress tensor and the charge current solve an old
puzzle raised in \cite{Bhattacharyya:2007vs} regarding the
fluid-gravity correspondence in large rotating AdS charged black
holes. The discrepancy found by the authors of
\cite{Bhattacharyya:2007vs} between the fluid dynamical predictions
and the known thermodynamics of rotating black holes is basically
resolved, once the qualitatively new effects due to the bulk
Chern-Simons interaction on hydrodynamics is taken into account. We
will reserve a deeper analysis of this issue along with a detailed
comparison of our second order fluid dynamical metric and gauge field
with charged black hole solutions to future work.

Though we have not yet worked out explicitly the position of the event
horizon in our metric solutions, it is plausible that the analysis of
\cite{Bhattacharyya:2008xc} can be easily extended to the case of
metrics with a global conserved charge(at least for the non-extremal
case).  This expectation, however has to be confirmed by an explicit
computation.  Further, it would be interesting to derive an entropy
current for the charged fluid following the proposal outlined in
\cite{Bhattacharyya:2008xc}.

In the bulk of our work we have refrained from commenting on the fluids
near extremality. The
preliminary analysis in \cite{Bhattacharyya:2007vs} suggests that
hydrodynamics would be
a valid description at least for some class of extremal solutions.
Unfortunately, we have not
been able to shed more light on this issue in the present work - the
perturbative metric
and the gauge field we have obtained are ill-behaved at the horizon in
the near-extremal limit.
In fact, even in the non-extremal case considered by us here, the
components of our solution
diverge at the inner horizon (this divergence is however shielded by
the outer horizon). We leave
unanswered the question of whether this is a co-ordinate artifact,
since this is a question which
requires a  detailed  analysis of the causal structure of our
solutions. If this divergence is physical
in the extremal case, it might have very interesting implications.
This is indeed one of the most
pressing questions opened up by the present work and it would be
interesting to understand the
extremal limit of our solutions more clearly.

\vspace{1cm}
%
\noindent
{\bf \large{Acknowledgement}}\\
%

The authors would like to thank Shiraz Minwalla for suggesting this
problem and providing guidance throughout the project. We would also
like to thank Veronica Hubeny, Mukund Rangamani and Sandip Trivedi for
helpful discussions during the project.  NB and SD would like to
thank D. Astefanesei, R. Gopakumar, D. Jatkar and A. Sen for
discussions. We also thank all the
students of Theory Physics student's room in TIFR for help and several
useful discussions. We would like to acknowledge the organizers of the
Monsoon Workshop on String theory at the Tata Institute of Fundamental
Research, Mumbai(organized by the International centre for Theoretical
Sciences) during which a major part of this work was done. NB, JB and
SD would like to thank the organisers of AdS/QCD school, 2008 at ICTP,
where the part of the work was done. PS would
like to thank the Tata Institute of Fundamental Research for its
hospitality and for partial support during the completion of this
work. PS was supported by Polish Ministry of Science and Information
Technologies grant 1P03B04029 (2005-2008).NB, SB, JB, SD and RL would
like to thank people of India for their
generous support to fundamental research. 
%

\appendix

\section{Charged conformal fluids and Weyl covariance} \label{app:weylcov}

Consider the hydrodynamic limit of a $3+1$ dimensional CFT with
one global conserved charge. The Weyl covariance of the CFT translates
into the Weyl covariance of its hydrodynamics. In turn, this implies that
the metric dual to fluid configurations of the CFT under consideration 
should also be invariant under boundary Weyl-transformations \cite{Loganayagam:2008is,Bhattacharyya:2008xc,Bhattacharyya:2008ji}.

In this section, we use the manifestly Weyl-covariant formalism introduced
in \cite{Loganayagam:2008is} to examine the constraints that Weyl-covariance
imposes on the conformal hydrodynamics and its metric dual. We begin by 
introducing a Weyl-covariant derivative acting on a general tensor field $Q^{\mu\ldots}_{\nu\ldots}$ with weight $w$ (by which we mean that the tensor
field transforms as $Q^{\mu\ldots}_{\nu\ldots}=e^{-w\phi}\tilde{Q}^{\mu\ldots}_{\nu\ldots}$
under a Weyl transformation of the boundary metric  $g_{\mu\nu}=e^{2\phi}g_{\mu\nu}$)
\begin{equation}\label{D:eq}
\begin{split}
\mathcal{D}_\lambda\ Q^{\mu\ldots}_{\nu\ldots} &\equiv \nabla_\lambda\ Q^{\mu\ldots}_{\nu\ldots} + w\  \mathcal{A}_{\lambda} Q^{\mu\ldots}_{\nu\ldots} \\ 
&+\left[{g}_{\lambda\alpha}\mathcal{A}^{\mu} - \delta^{\mu}_{\lambda}\mathcal{A}_\alpha  - \delta^{\mu}_{\alpha}\mathcal{A}_{\lambda}\right] Q^{\alpha\ldots}_{\nu\ldots} + \ldots\\
&-\left[{g}_{\lambda\nu}\mathcal{A}^{\alpha} - \delta^{\alpha}_{\lambda}\mathcal{A}_\nu  - \delta^{\alpha}_{\nu}\mathcal{A}_{\lambda}\right]  Q^{\mu\ldots}_{\alpha\ldots} - \ldots
\end{split}
\end{equation}
where the Weyl-connection $\mathcal{A}_\mu$ is related to
the fluid velocity $u^\mu$ via the relation
\begin{equation}\label{defA:eq}
\begin{split}
\mathcal{A}_\mu = u^\lambda\nabla_\lambda u_\mu - \frac{\nabla_\lambda u^\lambda}{3} {u}_\mu
\end{split}
\end{equation}

We can now use this Weyl-covariant derivative to enumerate all the Weyl-covariant
scalars, transverse vectors (i.e, vectors that are everywhere orthogonal to the 
fluid velocity field $u^\mu$) and the transverse traceless tensors in the charged hydrodynamics that involve no more than second order derivatives. We will do this enumeration `on-shell', i.e., we will enumerate those quantities which remain 
linearly independent even after the equations of motion are taken into account.
Our discussion here will closely parallel the discussion in section 4.1
of \cite{Bhattacharyya:2008ji} where a similar question was answered in the context of 
uncharged hydrodynamics coupled to a scalar with weight zero. However, we will use
a slightly different basis of Weyl-covariant tensors which is more suited for purposes
of this paper.

The basic fields in the charged hydrodynamics are the fluid velocity $u^\mu$ with
weight unity, the fluid temperature $T$ with with weight unity and the chemical 
potential $\mu$ with weight unity. This implies that an arbitrary function of 
$\mu/T$ is Weyl-invariant and hence one could always multiply a Weyl-covariant
tensor by such a function to get another Weyl-covariant tensor. Hence, in the
following list only linearly independent fields appear. To make contact with the 
conventional literature on hydrodynamics we will work with the charge density $n$
(with weight $3$) rather than the chemical potential $\mu$. 

At one derivative level, there are no Weyl invariant scalars or pseudo-scalars. The only Weyl invariant transverse vector is $n^{-1}P^\nu_\mu \mathcal{D}_\nu n $. Finally, the  only Weyl-invariant transverse pseudo-vector $l_\mu$ and only one Weyl-invariant symmetric traceless transverse tensor $T\sigma_{\mu\nu}$.

At the two derivative level, there are five independent Weyl-invariant scalars\footnote{We shall follow the notations of \cite{Loganayagam:2008is} in the rest of this section(except for the curvature tensors which differ by a sign from the curvature tensors in \cite{Loganayagam:2008is}. In particular, we recall the following definitions 
\begin{equation}\label{weylcov:eq}
\begin{split}
\mathcal{R} = R +6 \nabla_\lambda \mathcal{A}^\lambda - 6 \mathcal{A}_\lambda \mathcal{A}^\lambda \ ; \qquad& \mathcal{D}_\mu u_\nu = \sigma_{\mu\nu} + \omega_{\mu\nu} \\
\mathcal{D}_\lambda \sigma^{\mu\lambda} = \nabla_\lambda \sigma^{\mu\lambda}- 3 \mathcal{A}_\lambda \sigma^{\mu\lambda} \ ;\qquad& \mathcal{D}_\lambda \omega^{\mu\lambda} = \nabla_\lambda \omega^{\mu\lambda}- \mathcal{A}_\lambda \omega^{\mu\lambda}  \\
\end{split}
\end{equation} 
Note that in a flat space-time, $R$ is zero but $\mathcal{R}$ is not.} 
\begin{equation} \label{scalweyl}
\begin{split}
T^{-2}\sigma_{\mu\nu}\sigma^{\mu\nu},\quad T^{-2}\omega_{\mu\nu}\omega^{\mu\nu},\quad  T^{-2}\mathcal{R}, \quad T^{-2}n^{-1}P^{\mu\nu}\mathcal{D}_\mu \mathcal{D}_\nu n \quad  \text{and} \quad T^{-2}n^{-2}P^{\mu\nu}\mathcal{D}_\mu n \mathcal{D}_\nu n 
\end{split}
\end{equation}
one Weyl-invariant pseudo-scalar $T^{-2}n^{-1}l^{\mu}\mathcal{D}_\mu n$ and four independent Weyl-invariant transverse vectors  
\begin{equation}\label{vectweyl}
\begin{split}
T^{-1}P_\mu^{\nu}\mathcal{D}_\lambda\sigma_{\nu}{}^\lambda, \qquad T^{-1}P_\mu^{\nu}\mathcal{D}_\lambda\omega_{\nu}{}^\lambda, \qquad  T^{-1}n^{-1}\sigma_\mu{}^{\lambda}  \mathcal{D}_\lambda n\qquad
\text{and} \qquad  T^{-1}n^{-1}\omega_\mu{}^{\lambda}  \mathcal{D}_\lambda n 
\end{split}
\end{equation}
and one Weyl-invariant transverse pseudo-vector $T^{-1}\sigma_{\mu\nu}\ l^\nu$.

There are eight Weyl-invariant symmetric traceless transverse tensors - 
\begin{equation} \label{symweyl}
\begin{split}
u^\lambda\mathcal{D}_\lambda\sigma_{\mu\nu}, &\quad  \omega_{\mu}{}^{\lambda}\sigma_{\lambda\nu}+\omega_{\nu}{}^{\lambda}\sigma_{\lambda\mu},
\quad \sigma_{\mu}{}^{\lambda}\sigma_{\lambda\nu}-\frac{P_{\mu\nu}}{3}\  \sigma_{\alpha\beta}\sigma^{\alpha\beta},\quad 
\omega_{\mu}{}^{\lambda}\omega_{\lambda\nu}+\frac{P_{\mu\nu}}{3}\  \omega_{\alpha\beta}\omega^{\alpha\beta},\\
n^{-1}\ \Pi_{\mu\nu}^{\alpha\beta}\
 \mathcal{D}_\alpha \mathcal{D}_\beta n ,&\quad 
n^{-2}\ \Pi_{\mu\nu}^{\alpha\beta}\ \mathcal{D}_\alpha n\ \mathcal{D}_\beta n, 
\quad C_{\mu\alpha\nu\beta}u^\alpha u^\beta \quad\text{and}\quad \frac{1}{4}\ \epsilon^{\alpha\beta\lambda}{}_{\mu}\ \epsilon^{\gamma\theta\sigma}{}_{\nu} C_{\alpha\beta\gamma\theta}\ u_\lambda u_{\sigma}.
\end{split}
\end{equation}
where we have introduced the projection tensor $\Pi_{\mu\nu}^{\alpha\beta}$ which projects out the transverse traceless symmetric part of  second rank tensors
\[ \Pi_{\mu\nu}^{\alpha\beta} \equiv \frac{1}{2}\left[P^\alpha_\mu P^\beta_\nu + P^\alpha_\nu P^\beta_\mu - \frac{2}{3} P^{\alpha\beta}P_{\mu\nu} \right] \]
and $C_{\mu\nu\alpha\beta}$ is the boundary Weyl curvature tensor. Further, there are four Weyl-invariant symmetric traceless transverse pseudo-tensors
\begin{equation}\label{sympseudoweyl}
\begin{split}
\mathcal{D}_{(\mu}l_{\nu)},\quad n^{-1} \Pi_{\mu\nu}^{\alpha\beta}l_\alpha \mathcal{D}_\beta n,   
&\quad n^{-1}\epsilon^{\alpha\beta\lambda}{}_{(\mu} \sigma_{\nu)\lambda}u_\alpha\mathcal{D}_\beta n
\quad\text{and}\quad \frac{1}{2} \epsilon_{\alpha\beta\lambda(\mu}C^{\alpha\beta}{}_{\nu)\sigma}u^\lambda u^\sigma .
\end{split}
\end{equation}

We will now restrict ourselves to the case where the boundary metric is flat. In this
case the last two tensors appearing in  \eqref{symweyl} and the last tensor appearing
in \eqref{sympseudoweyl} are identically zero whereas, contrary to what one might
naively expect, the Weyl-covariantised Ricci scalar $\mathcal{R}$ would still 
be non-zero.

We will now relate the rest of the Weyl-covariant scalars, transverse vectors and
symmetric, traceless transverse tensors listed above to the quantities 
appearing in the  table~\ref{tablevtdefs}. 

There are six scalar/pseudo-scalar Weyl covariant combinations given by 
\begin{equation}
\begin{split}
W_{s}^{1} &\equiv \sigma_{\mu\nu}\sigma^{\mu\nu} = \text{ST5} \\
W_{s}^{2} &\equiv \omega_{\mu\nu}\omega^{\mu\nu} =\frac{1}{2}\text{ST4}\\
W_{s}^{3} &\equiv \mathcal{R} = 14\ \text{ST1}+\frac{2}{3} \text{ST3}-\text{ST4}+ 2\text{ST5}-\frac{\text{S3}}{m} \\
W_{s}^{4} &\equiv n^{-1} P^{\mu\nu}\mathcal{D}_\mu \mathcal{D}_\nu n =\frac{1}{q}\left[ \text{QS2}-\frac{3 q}{4 m} \text{S3} + 18 \text{q} \text{ST1} + 5 \text{QS5}\right] \\
W_{s}^{5} &\equiv n^{-2} P^{\mu\nu}\mathcal{D}_\mu n \ \mathcal{D}_\nu n = \frac{1}{q^2}\left[\text{QS4}+6 \text{q} \text{QS5} + 9 \text{q}^2 \text{ST1}\right] \\
W_{s}^{6} &\equiv l^{\mu}\mathcal{D}_\mu q = \text{QS3}+ 3 \text{q} \text{ST2}.\\
\end{split}
\end{equation}
and five vector/pseudo-vector Weyl covariant combinations given by 
\begin{equation}
\begin{split}
(W_v)^{1}_{\mu} &\equiv P_{\mu}^\nu \mathcal{D}_\lambda \sigma_{\nu}{}^\lambda = \frac{5 \text{V4}}{9}+\frac{5 \text{V5}}{9}+\frac{5
   \text{VT1}}{3}-\frac{5 \text{VT2}}{12}-\frac{11
   \text{VT3}}{6} \\
(W_v)^{2}_{\mu} &\equiv P_{\mu}^\nu \mathcal{D}_\lambda \omega_{\nu}{}^\lambda =\frac{5
   \text{V4}}{3}-\frac{\text{V5}}{3}-\text{VT1}-\frac{
   \text{VT2}}{4}+\frac{\text{VT3}}{2} \\
(W_v)^{3}_{\mu} &\equiv l^\lambda\sigma_{\mu\lambda}  = \text{VT5}\\
(W_v)^{4}_{\mu} &\equiv n^{-1} \sigma_{\mu}{}^\lambda\mathcal{D}_\lambda n = \frac{1}{q}\left[\text{QV4}+3 \text{q} \text{VT3}\right] \\
(W_v)^{5}_{\mu} &\equiv n^{-1}\omega_{\mu}{}^\lambda\mathcal{D}_\lambda n = \frac{1}{2q}\left[\text{QV3}+3 \text{q} \text{VT2}\right] \\
\end{split}
\end{equation}

In the tensor sector, there are nine Weyl-covariant combinations 
\begin{equation}\label{weyltendef}
\begin{split}
WT^{(1)}_{\mu\nu} &= u^\lambda \mathcal{D}_{\lambda} \sigma_{\mu \nu} = TT1 + \frac{1}{3} TT4 + T3 .\\
WT^{(2)}_{\mu\nu} &= -2 \left( \omega^{\mu}{}_{\lambda} \sigma^{\lambda \nu} 
+ \omega^{\nu}{}_{\lambda} \sigma^{\lambda \mu} \right) = TT7 .\\
WT^{(3)}_{\mu\nu} &= \sigma^{\mu}{}_{\lambda} \sigma_{\lambda \nu} - \frac{1}{3} P^{\mu \nu} 
\sigma^{\alpha \beta} \sigma_{\alpha \beta} = TT6.\\
WT^{(4)}_{\mu\nu} &= 4 \left( \omega^{\mu}{}_{\lambda} \omega_{\lambda \nu} + \frac{1}{3} P^{\mu \nu} 
\omega^{\alpha \beta} \omega_{\alpha \beta} \right) =  TT5 .\\
WT^{(5)}_{\mu\nu} &= n^{-1}\Pi_{\mu\nu}^{\alpha\beta} \mathcal{D}_{\alpha} \mathcal{D}_{\beta} n\\
&= \frac{1}{q}\left[QT1 +8 QT4+ 15 q TT1 +q TT4 +3 q T3 +3 q TT6+\frac{3 q}{4} TT5\right] \\
WT^{(6)}_{\mu\nu} &= n^{-2}\Pi_{\mu\nu}^{\alpha\beta} \mathcal{D}_{\alpha} n \mathcal{D}_{\beta} n
= \frac{1}{q^2}\left[QT3 + 6 q QT4 + 9 q^2 TT1 \right]\\
WT^{(7)}_{\mu\nu} &= \mathcal{D}_\mu l_{\nu} + \mathcal{D}_\nu l_{\mu} = 4 TT2 + 2 T2 - TT3.\\
WT^{(8)}_{\mu\nu} &= n^{-1} \Pi_{\mu\nu}^{\alpha\beta}l_\alpha \mathcal{D}_\beta n = \frac{1}{q}\left[QT2 + 3\  q\  TT2\right].\\
WT^{(9)}_{\mu\nu} &= n^{-1}\epsilon^{\alpha\beta\lambda}{}_{(\mu} \sigma_{\nu)\lambda}u_\alpha\mathcal{D}_\beta n =\frac{1}{q}\left[QT5 - \frac{3}{2}\  q\  TT2+ \frac{3}{2}\  q\  TT3\right].
\end{split}
\end{equation}
%

\section{Source Terms in Scalar Sector: Second Order}\label{appscl2}
There are three source terms in scalar sector at second order $S_k(r)$, $S_h(r)$ and
$S_M(r)$. They are quite complicated functions. Here we provide
the explicit form of these source terms in terms of weyl covariant quantities.

The source term $S_k$ is given by
\ben
S_C = \sum_{i=1}^6 s_i^{(C)} W_s^{i}.
\een
The Weyl covariant terms $W_s^{i}$ are given in \S \ref{app:weylcov}.
The functions $s_i^{(k)}$s are given by,
\begin{equation}
 \begin{split}
s_1^{(C)} &= \frac{r \left(4 \left(m_0-3 r^4\right) \left(r^2+r
   R+R^2\right)
   F_2\left(\frac{r}{R},\frac{m_0}{R
   ^4}\right)+R \left(m_0 (r+R)-2
   R^3 \left(r^2+r
   R+R^2\right)\right)\right)}{3 R
   (r+R) \left(-m_0+r^4+r^2
   R^2+R^4\right)}\\
s_2^{(C)} &= \frac{1}{3
   m_0^2 r^7}\left. \Big( -m_0^3 \left(r^4+2 r^2 R^2+36 R^4
   \kappa ^2\right)+2 m_0^2 \left(18 r^4 R^4
   \kappa ^2+r^2 R^6+36 R^8 \kappa ^2\right) \right.\\ &\left. -36
   m_0 R^8 \kappa ^2 \left(2
   r^4+R^4\right)+36 r^4 R^{12} \kappa ^2 \right. \Big)\\
s_3^{(C)} &= \frac{r}{3} \\
s_4^{(C)} &= \frac{2 r^2 \left(m_0-R^4\right) \left(r
   F_1^{(1,0)}\left(\frac{r}{R},\frac{m_0}{
   R^4}\right)+6 R
   F_1\left(\frac{r}{R},\frac{m_0}{R
   ^4}\right)\right)}{R^6} \\
s_5^{(C)} & = -\frac{1}{2 R^{16} \left(m_0-3
   R^4\right)} \left( r^2 \left(m_0-R^4\right) \left(24
   R^4
   F_1\left(\frac{r}{R},\frac{m_0}{R
   ^4}\right) \left(r^3 \left(m_0^2-4 m_0
   R^4   \right. \right. \right. \right. \\ &+ \left. \left. \left. \left. 3 R^8\right)
   F_1^{(2,0)}\left(\frac{r}{R},\frac{m_0}{
   R^4}\right)+11 r^2 R \left(m_0^2-4
   m_0 R^4+3 R^8\right)
   F_1^{(1,0)}\left(\frac{r}{R},\frac{m_0}{
   R^4}\right)\right. \right. \right. \\ &+ \left. \left. \left. 6 m_0 R^7-4
   R^{11}\right)+r \left(r^2 R^2
   \left(m_0^2 \left(25 r^2-13
   R^2\right)+m_0 \left(-25 r^6-75 r^2
   R^4+52 R^6\right)\right. \right. \right. \right. \\ &+ \left. \left. \left. \left. 75 r^6 R^4-39
   R^{10}\right)
   F_1^{(1,0)}\left(\frac{r}{R},\frac{m_0}{
   R^4}\right)^2+r
   F_1^{(2,0)}\left(\frac{r}{R},\frac{m_0}{
   R^4}\right) \left(4 R^9
   \left(m_0-R^4\right)\right. \right. \right. \right. \\ &- \left. \left. \left. \left.r^3 \left(m_0^2
   \left(R^2-r^2\right)+m_0 \left(r^6+3 r^2
   R^4-4 R^6\right)+3 R^4
   \left(R^6-r^6\right)\right)
   F_1^{(2,0)}\left(\frac{r}{R},\frac{m_0}{
   R^4}\right)\right)\right. \right. \right. \\ &+ \left. \left. \left. 2 R
   F_1^{(1,0)}\left(\frac{r}{R},\frac{m_0}{
   R^4}\right) \left(-5 r^3 \left(m_0^2
   \left(R^2-r^2\right)+m_0 \left(r^6+3 r^2
   R^4-4 R^6\right)\right. \right.\right.\right. \right. \\ &+ \left. \left. \left.\left. \left. 3 R^4
   \left(R^6-r^6\right)\right)
   F_1^{(2,0)}\left(\frac{r}{R},\frac{m_0}{
   R^4}\right)+26 m_0 R^9-22
   R^{13}\right)\right.\right. \right. \\ &+ \left. \left. \left.16 m_0 R^6
   \left(m_0-R^4\right)
   F_1^{(1,1)}\left(\frac{r}{R},\frac{m_0}{
   R^4}\right)\right)+96 m_0 R^7
   \left(m_0-R^4\right)
   F_1^{(0,1)}\left(\frac{r}{R},\frac{m_0}{
   R^4}\right)\right. \right. \\ &+ \left. \left. 288 r R^6 \left(m_0-3
   R^4\right) \left(m_0-R^4\right)
   F_1\left(\frac{r}{R},\frac{m_0}{R
   ^4}\right)^2\right) \right)\\
s_6^{(C)} &= \frac{2 \sqrt{3} \kappa  \left(m_0-r^4\right)
   \left(R^4-m_0\right) \left(5 R
   F_1^{(1,0)}\left(\frac{r}{R},\frac{m_0}{
   R^4}\right)+r
   F_1^{(2,0)}\left(\frac{r}{R},\frac{m_0}{
   R^4}\right)\right)}{m_0 R^7}.
 \end{split}
\end{equation}
The source term $S_h$ is given by
\ben
S_h = \sum_{i=1}^6 s_i^{(h)} W_s^{i},
\een
where the functions $s_i^{(h)}$'s are given by
\begin{equation}
\begin{split}
s_1^{(h)} &= \frac{1}{3 R
   (r+R)^2 \left(-m_0+r^4+r^2
   R^2+R^4\right)^2} \left( 2 r \left(2 \left(m_0 \left(4 r^3+8 r^2 R+6
   r R^2+3 R^3\right)\right. \right. \right. \\ & \left. \left. \left.-3 R^3 \left(r^2+r
   R+R^2\right)^2\right)
  F_2\left(\frac{r}{R},\frac{m_0}{R
   ^4}\right)+r^2 R \left(r^2+r
   R+R^2\right)^2\right) \right), \\
s_2^{(h)} &= \frac{2}{3
   r^7} \left(r^4-\frac{36 R^4 \kappa ^2
   \left(m_0-R^4\right)^2}{m_0^2}\right), \\
s_3^{(h)} &=0,\\
s_4^{(h)} &=0,\\
s_5^{(h)} &= \frac{r^7 \left(R^4-m_0\right)}{R^{16}} \left(5 R
   F_1^{(1,0)}\left(\frac{r}{R},\frac{m_0}{
   R^4}\right)+r
   F_1^{(2,0)}\left(\frac{r}{R},\frac{m_0}{
   R^4}\right)\right)^2\\
s_6^{(h)} &= \frac{4 \sqrt{3} \kappa  \left(R^4-m_0\right)}{m_0 R^7}
   \left(5 R
   F_1^{(1,0)}\left(\frac{r}{R},\frac{m_0}{
   R^4}\right)+r
   F_1^{(2,0)}\left(\frac{r}{R},\frac{m_0}{
   R^4}\right)\right).
\end{split}
\end{equation}

Finally the source term $S_M(r)$ is given by
\ben
S_M(r) = \sum_{i=1}^6 s_i^{(M)} W_s^{i}\ ,
\een
with the functions $s_i^{(M)}$ being given by
\begin{equation}
\begin{split}
s_1^{(M)} &= \frac{4 r \sqrt{R^2 \left(m_0-R^4\right)}
   \left(r^2+r R+R^2\right)
  F_2\left(\frac{r}{R},\frac{m_0}{R
   ^4}\right)}{R (r+R) \left(-m_0+r^4+r^2
   R^2+R^4\right)} \\
s_2^{(M)} &=-\frac{2 \sqrt{R^2 \left(m_0-R^4\right)}
   \left(m_0^2 r^4+12 R^2 \kappa ^2
   \left(m_0-R^4\right) \left(2 m_0 r^2+3
   m_0 R^2-3
   R^6\right)\right)}{m_0^2 r^7} \\
s_3^{(M)} &= 0 \\
s_4^{(M)} &=-\frac{r^5 \sqrt{R^2 \left(m_0-R^4\right)}
   \left(5 R
   F_1^{(1,0)}\left(\frac{r}{R},\frac{m_0}{
   R^4}\right)+r
   F_1^{(2,0)}\left(\frac{r}{R},\frac{m_0}{
   R^4}\right)\right)}{R^9}\\
s_5^{(M)} &=\frac{r^5 \left(R^2
   \left(m_0-R^4\right)\right)^{3/2} }{R^{17}
   \left(3 R^4-m_0\right)} \left(r^2
   \left(-\left(6 r \left(m_0-3 R^4\right)
   F_1\left(\frac{r}{R},\frac{m_0}{R
   ^4}\right)+R^5\right)\right)
   F_1^{(3,0)}\left(\frac{r}{R},\frac{m_0}{
   R^4}\right) \right. \\ & \left. -2 \left(15 r^2 R \left(m_0-3
   R^4\right)
   F_1^{(1,0)}\left(\frac{r}{R},\frac{m_0}{
   R^4}\right)^2+\left(3 r \left(m_0-3
   R^4\right) \left(r^2
   F_1^{(2,0)}\left(\frac{r}{R},\frac{m_0}{
   R^4}\right)\right.\right.\right.\right. \\ & \left.\left.\left.\left.+35 R^2
   F_1\left(\frac{r}{R},\frac{m_0}{R
   ^4}\right)\right)+20 R^7\right)
   F_1^{(1,0)}\left(\frac{r}{R},\frac{m_0}{
   R^4}\right)+R \left(2 m_0 r R
   F_1^{(2,1)}\left(\frac{r}{R},\frac{m_0}{
   R^4}\right)\right.\right.\right. \\ & \left.\left.\left.+r \left(39 r \left(m_0-3
   R^4\right)
   F_1\left(\frac{r}{R},\frac{m_0}{R
   ^4}\right) +7 R^5\right)
   F_1^{(2,0)}\left(\frac{r}{R},\frac{m_0}{
   R^4}\right) \right. \right. \right. \\ & \left. \left. \left.  +10 m_0 R^2
   F_1^{(1,1)}\left(\frac{r}{R},\frac{m_0}{
   R^4}\right)\right)\right)\right) \\
s_6^{(M)} &= \frac{\sqrt{3} \kappa  \left(R^4-m_0\right) }{m_0 r^2
   R^7 \sqrt{R^2
   \left(m_0-R^4\right)}}
   \left(-m_0 r^4 R
   F_1^{(3,0)}\left(\frac{r}{R},\frac{m_0}{
   R^4}\right)+r^4 R^5
   F_1^{(3,0)}\left(\frac{r}{R},\frac{m_0}{
   R^4}\right)\right. \\ &\left.+r^2 R \left(20 m_0 r^2-17
   m_0 R^2+17 R^6\right)
   F_1^{(1,0)}\left(\frac{r}{R},\frac{m_0}{
   R^4}\right)\right. \\ &\left.+r^3 \left(4 m_0 r^2-7 m_0
   R^2+7 R^6\right)
   F_1^{(2,0)}\left(\frac{r}{R},\frac{m_0}{
   R^4}\right)+4 R^9\right) \\
\end{split}
\end{equation}
%

\section{Source Terms in Vector Sector: Second Order} \label{appvec2}

The source term in the vector sector at second order $S_E^{\text{vec}}(r)$ in \eqref{vecdyn1} is given by
\ben
(S_E^{\text{vec}})_i(r) = \sum_{l=1}^5 r_{l}^{(E)} (W_v)^l_i
\een
where the Weyl covariant quantities $W_v^i$'s are given in Appendix \ref{app:weylcov}
and the functions $s_{i}^{(E)}$ are given by
\begin{equation}
\begin{split}
r_{1}^{(E)} &= \frac{r^2+r R+R^2}{3 (r+R)
   \left(-m_0+r^4+r^2 R^2+R^4\right)},\\
r_{2}^{(E)} &= \frac{1}{3 r^3},\\
r_{3}^{(E)} &= \frac{\kappa  \left(R^2
   \left(m_0-R^4\right)\right)^{3/2}
   \left(m_0 (r+2 R)+3 r \left(r^2+r
   R+R^2\right)^2\right)}{\sqrt{3} m_0 r^3
   (r+R)^2 \left(-m_0+r^4+r^2
   R^2+R^4\right)^2},\\ 
r_{4}^{(E)} &=\frac{\left(m_0-R^4\right)}{3 R^6 (r+R)^2
   \left(-m_0+r^4+r^2 R^2+R^4\right)^2} \Bigg( -6 r^2
   (r+R) \left(r^2+r R+R^2\right)
   \left(-m_0  \right. \\ & \left.  + r^4+r^2 R^2+R^4\right)
   F_1^{(1,0)}\left(\frac{r}{R},\frac{m_0}{
   R^4}\right)-6 r R \left(3 \left(r^2+r
   R+R^2\right)^2 \left(r^3+2
   R^3\right) \right. \\ & \left.  -m_0 \left(7 r^3+14 r^2 R+12
   r R^2+6 R^3\right)\right)
   F_1\left(\frac{r}{R},\frac{m_0}{R^4}
   \right) \\ &  -\frac{R^8 \left(m_0 (2
   r+R)+3 R \left(r^2+r
   R+R^2\right)^2\right) }{m_0-3
   R^4}\Bigg),\\
r_{5}^{(E)} &= \frac{\left(R^4-m_0\right) \left(r \left(9 R
  F_1^{(1,0)}\left(\frac{r}{R},\frac{m_0}{R^4}\right)+r
  F_1^{(2,0)}\left(\frac{r}{R},\frac{m_0}{R^4}\right)\right)+6
  R^2 F_1\left(\frac{r}{R},\frac{m_0}{R^4}\right)\right)}{r^2
  R^7}.\\
\end{split}
\end{equation}

The other source term in the vector sector at second order $S_{M}^{\text{vec}}(r)$ in \eqref{vecdyn2} 
is given by
\ben
(S_M^{\text{vec}})_i(r) = \sum_{l=1}^5 r_{l}^{(M)} (W_v)^l_i,
\een
where the coefficient functions $r_{i}^{(M)}$ are given by
\begin{equation}
\begin{split}
r_{1}^{(M)} &= 0, \\
r_{2}^{(M)} &=\frac{2 \sqrt{3} \sqrt{R^2 \left(m_0-R^4\right)} \left(m_0 r^2+24
   R^2 \kappa ^2 \left(R^4-m_0\right)\right)}{m_0 r^5}, \\
r_{3}^{(M)} &= \frac{6 R \kappa  \left(m_0-R^4\right) }{m_0
   r^5 (r+R) \left(-m_0+r^4+r^2 R^2+R^4\right)}\Bigg(r^2 R \left(r
   \left(r^2+r R+R^2\right) \left(3 r^3+R^3\right) \right. \\ & \left. -m_0 \left(3
   r^2+3 r R+2 R^2\right)\right)-8 m_0 (r+R)
   \left(-m_0+r^4+r^2 R^2+R^4\right)
   F_2\left(\frac{r}{R},\frac{m_0}{R^4}\right)\Bigg),\\ 
r_{4}^{(M)} &=-\frac{2 \sqrt{3} \sqrt{R^2 \left(m_0-R^4\right)}}{R^6 (r+R)
   \left(-m_0+r^4+r^2 R^2+R^4\right)} \Bigg(r^2
   (r+R) \left(-m_0+r^4 \right. \\ & \left. +r^2 R^2+R^4\right) \left(5 R
   F_1^{(1,0)}\left(\frac{r}{R},\frac{m_0}{R^4}\right)+r
   F_1^{(2,0)}\left(\frac{r}{R},\frac{m_0}{R^4}\right)\right) \\ & +12
   r R \left(m_0-R^4\right) \left(r^2+r R+R^2\right)
   F_1\left(\frac{r}{R},\frac{m_0}{R^4}\right)+R^6
   \left(r^2+r R+R^2\right)\Bigg), \\
r_{5}^{(M)} &= \frac{2 \sqrt{3} \sqrt{R^2 \left(m_0-R^4\right)}}{m_0 r^5 R^6} \Bigg(6 m_0 r^3
   \left(m_0-R^4\right) \left(R
   F_1\left(\frac{r}{R},\frac{m_0}{R^4}\right)-r
   F_1^{(1,0)}\left(\frac{r}{R},\frac{m_0}{R^4}\right)\right) 
    \\ & +m_0 r^2 R^6+24 R^8 \kappa ^2
   \left(R^4-m_0\right)\Bigg).\\
\end{split}
\end{equation}
%

\section{Source Terms in Tensor Sector: Second Order} \label{appten2}

In this appendix we provide the source of the dynamical equation \eqref{dyneqten}.
We report the result in terms of the parameters $M$ and $R$ and the variable $\rho$
defined in \eqref{rhoMQ:eq}. The source ${{\bf T}_{ij}(\rho)}$ in  \eqref{dyneqten} is given by
\begin{equation}
\begin{split}
{\bf T}_{ij}(r) = \sum_{l=1}^9 \tau_{l}(r) ~WT^{(l)}_{ij},
\end{split}
\end{equation}
where the weyl-covariant terms $WT^{(l)}_{ij}$ 
are defined in Appendix \ref{app:weylcov} 
in equation \eqref{weyltendef}.
The coefficient of the weyl-covariant terms 
in the above source is given by
\begin{equation}
\begin{split}
{\tau}_1(r) = & ~ \frac{3 r
   F_2\left(\frac{r}{R},
   \frac{m_0}{R^4}\right)}{R}+\frac{m_0
   (r+R)-\left(r^2+r R+R^2\right)
   \left(3 r^3+R^3\right)}{(r+R)
   \left(-m_0+r^4+r^2 R^2+R^4\right)},\\
{\tau}_2(r) = & ~ -\frac{1}{2 R} \Bigg( 3 r
   F_2\left(\frac{r}{R},
   \frac{m_0}{R^4}\right)-\frac{2 r^3 R \left(r^2+r
   R+R^2\right)}{(r+R)
   \left(-m_0+r^4+r^2
   R^2+R^4\right)}+R \Bigg),\\
{\tau}_3(r) = & ~ \frac{6 r
   F_2\left(\frac{r}{R},
   \frac{m_0}{R^4}\right)}{R}+\frac{2 \left(m_0
   (r+R)-2 r^3 \left(r^2+r
   R+R^2\right)\right)}{(r+R)
   \left(-m_0+r^4+r^2 R^2+R^4\right)},\\
{\tau}_4(r) = & ~ \frac{18 R^4 \kappa ^2
   \left(m_0-R^4\right)^2 \left(-m_0 r^2+4
   m_0 R^2+r^6-4 R^6\right)}{m_0^2
   r^{10}}-\frac{m_0 r^2+2 m_0 R^2+r^6-2
   R^6}{2 r^6},\\
{\tau}_5(r) = & ~ \frac{1}{R^6} \Bigg( 6 r \left(R^4-m_0\right) \left(r
   F_1^{(1,0)}\left(\frac{r}{R},\frac{m_0}
   {R^4}\right)+3 R
   F_1\left(\frac{r}{R},
   \frac{m_0}{R^4}\right)\right) \Bigg),\\
{\tau}_6(r) = & ~ \frac{3 r \left(m_0-R^4\right) }{2 R^{16} \left(m_0-3
   R^4\right)} \Bigg(r
   \left(R^2
   F_1^{(1,0)}\left(\frac{r}{R},\frac{m_0}
   {R^4}\right) \left(r^2 \left(m_0-3
   R^4\right) \left(-25 m_0 r^2 
   \right. \right. \right. \\& \left. \left. \left. +37 m_0
   R^2+25 r^6-37 R^6\right)
   F_1^{(1,0)}\left(\frac{r}{R},\frac{m_0}
   {R^4}\right)+30 m_0 R^8-2
   R^{12}\right) \right. \\& \left. -r^4 \left(3
   R^4-m_0\right) (r-R) (r+R)
   \left(-m_0+r^4+r^2 R^2+R^4\right)
   F_1^{(2,0)}\left(\frac{r}{R},\frac{m_0}
   {R^4}\right)^2 \right. \\& \left. +2 r R \left(5 r^2 \left(3
   R^4-m_0\right) (R-r) (r+R)
   \left(-m_0+r^4 \right. \right. \right.\\& \left. \left. \left.+r^2 R^2+R^4\right)
   F_1^{(1,0)}\left(\frac{r}{R},\frac{m_0}
   {R^4}\right)+R^8
   \left(m_0+R^4\right)\right)
   F_1^{(2,0)}\left(\frac{r}{R},\frac{m_0}
   {R^4}\right) \right. \\& \left. +16 m_0 R^6
   \left(m_0-R^4\right)
   F_1^{(1,1)}\left(\frac{r}{R},\frac{m_0}
   {R^4}\right)\right)+48 m_0 R^7
   \left(m_0-R^4\right)
   F_1^{(0,1)}\left(\frac{r}{R},\frac{m_0}
   {R^4}\right) \\ & +24 R^{11} \left(3 m_0-2
   R^4\right)
   F_1\left(\frac{r}{R},\frac{m_0}{
   R^4}\right)\Bigg)\\
{\tau}_7(r) = & ~ \frac{3 \sqrt{3} \kappa  \left(R^2
   \left(m_0-R^4\right)\right)^{3/2}}{2
   m_0 r^5},\\
{\tau}_8(r) = & ~ \frac{3 \sqrt{3} \kappa  \left(R^2
   \left(m_0-R^4\right)\right)^{3/2} }{m_0
   r^5 R^9} \left(2 r^2
   \left(R \left(-m_0 \left(5
   r^2+R^2\right)+5 r^6+R^6\right)
   F_1^{(1,0)}\left(\frac{r}{R},\frac{m_0}
   {R^4}\right) \right. \right. \\ & \left. \left. +r \left(m_0
   \left(R^2-r^2\right)+r^6-R^6\right)
   F_1^{(2,0)}\left(\frac{r}{R},\frac{m_0}
   {R^4}\right)\right)+R^9\right),\\
{\tau}_9(\rho) = & ~ 0.
\end{split}
\end{equation}
%

\section{Comparison with Erdmenger et. al. \cite{Erdmenger:2008rm} } \label{map}

Firstly we shall present a dictionary of relations between the quantities defined in 
\cite{Erdmenger:2008rm} and those in this paper. To avoid confusion we shall use a 
subscript `$E$' to denote the quantities in \cite{Erdmenger:2008rm}.

The charge and mass of the black brane in the two papers are related by
\begin{equation}
\begin{split}
(Q)_E &= -q \\
(b)_E^4 &= \frac{1}{m}.
\end{split}
\end{equation}

Also the gauge field in \cite{Erdmenger:2008rm} is twice the gauge field in our paper
\begin{equation}
 (A_{\mu})_E = 2 A_{\mu}.
\end{equation}

We list the relation between several other quantities in the two papers
\begin{equation}
\begin{split}
(r_+)_E &= R \\
(r_-)_E &= R \sqrt{\left( Q^2 + \frac{1}{4}\right)^{\frac{1}{2}} - \frac{1}{2}} \\
\mu_E &= -\frac{\sqrt{3} q}{R^2} = -2 \mu \\
T_E &= \frac{R}{2 \pi} (3-M) = \frac{R}{2 \pi} (2-Q^2) \\
N_E^2 &= \frac{\pi}{2 G_5}\\
(\sigma_{\mu \nu})_E &= 2 \sigma_{\mu \nu}\\
(l_{\mu})_E &= - l_{\mu}.
\end{split}
\end{equation}

Finally the various functions that go into the first order metric and the gauge field are 
related by
\begin{equation}
\begin{split}
(F(r))_E &= \frac{1}{R} F_2(\rho,M) \\
(j^{\kappa}(r))_E &= \frac{\sqrt{3}Q(2-Q^2)^3}{2 \pi R (2+3Q^2+Q^4)} F_1(\rho,M) \\
(a^{\kappa}(r))_E &= - \frac{\rho^5 (2-Q^2)^3}{4 \pi (2+3Q^2+Q^4)} F_1^{(1,0)}(\rho,M).
\end{split}
\end{equation}

These statements are true only up to zeroth order in the 
expansion of R in terms of the boundary derivatives.
Further for the tensor sector matching we have to use the following relations
\begin{equation}
\begin{split}
\mathcal{D}_i \left( \frac{\mu}{T} \right) &= \frac{2 \pi \sqrt{3} (2+3Q^2 +Q^4)}{R^3 (2-Q^2)^3} \mathcal{D}_i q \\
\mathcal{D}_i \mathcal{D}_j \left( \frac{\mu}{T} \right) &= \frac{2 \pi \sqrt{3} (2+3Q^2 +Q^4)}{R^3 (2-Q^2)^3} 
                  \mathcal{D}_i \mathcal{D}_j q + \frac{2 \pi \sqrt{3} Q (1+Q^2)(60 + 40 Q^2 + Q^4)}{R^6 (2-Q^2)^5}
                  \mathcal{D}_i q \mathcal{D}_j q ,
\end{split}
\end{equation}
where, $$\mu = \frac{\sqrt{3} q}{2 R^2}; \quad \quad T = \frac{R}{2 \pi} (2-Q^2),$$
are respectively the chemical potential and the temperature in our notation.

Using this dictionary our stress tensor  and charge current mathches 
perfectly with \cite{Erdmenger:2008rm}.


\bibliography{chargedBH_FinalDraft}

\end{document}